\documentclass[aps,prb,showpacs,showkeys,twocolumn]{revtex4}
\usepackage{amsfonts}
\usepackage{amsmath}
\usepackage{amssymb}
\usepackage{epsf}
\usepackage{rotate}
\usepackage{bookmath}
\usepackage{color}
\usepackage{mycolors}
\usepackage{feynmp}
\usepackage{feynman_defs}
\epsfverbosetrue
\begin{document}
\title{Theory of Heat Transport of Normal Liquid $^3$He in Aerogel}
\author{J. A. Sauls}
\affiliation{Department of Physics \& Astronomy, Northwestern University, 
             Evanston, IL 60208}
\author{Priya Sharma}
\affiliation{Department of Physics, Royal Holloway University of London, 
             Egham, Surrey, TW20 0EX, UK}
\date{\today}
\begin{abstract}
The introduction of liquid \He\ into silica aerogel provides us a with model system in which to
study the effects of disorder on the properties of a strongly correlated Fermi liquid. The transport
of heat, mass and spin exhibits cross-over behavior from a high temperature regime, where inelastic
scattering dominates, to a low temperature regime dominated by elastic scattering off the aerogel.
We report exact and approximate solutions to the Boltzmann-Landau transport equation for the
thermal conductivity of liquid \He, including elastic scattering of quasiparticles by
the aerogel and inelastic quasiparticle collisions. 
These results provide quantitative predictions for the transport properties of liquid \He\ in
aerogel over a wide range of pressure, temperature and aerogel density. In particular, we obtain a
scaling function, $F(T/T_{\star})$, for the normalized thermal conductivity,
$\kappa/\kappa_{\text{el}}$, in terms of a reduced temperature, $T/T_{\star}$, where $T^{\star}$ is
a cross-over temperature defined by the elastic and inelastic collision rates.
Theoretical results are compared with the available experimental data for the thermal
conductivity.
\end{abstract}
\maketitle
\section {Introduction}

Aerogels are extremely low density solids formed as a rigid network of silica strands and
clusters having a typical diameter of 30-50 {\AA} and porosities ($\varrho$) above
99\%.\cite{fricke86} They turn out to be an ideal system for studying the effects of quenched
random disorder on the otherwise pure, ordered phases of liquid \He. When impregnated with
liquid \He, the aerogel is found to have dramatic effects on the transport properties and the
phase diagram of liquid \He, although basic thermodynamic features characteristic of the Fermi
liquid, such as the compressibility, magnetization and heat capacity, are essentially
unchanged.\cite{hal04}
The low temperature transport of mass, heat and magnetization is substantially
reduced.\cite{por95,ree02,sau05} In addition, the superfluid transition temperature as well as the
superfluid order parameter are strongly suppressed relative to their bulk values.\cite{por95,spr95}

In this paper we consider the effects of scattering of \He\ quasiparticles off a uniformly
distributed random potential representing the aerogel structure, referred to as the
``homogeneous scattering model'' (HSM).\cite{thu98} We obtain exact and approximate solutions to
the Boltzmann-Landau transport equation for the thermal conductivity of liquid \He,
including both inelastic collisions between quasiparticles and elastic scattering of
quasiparticles by the random potential.
The chief inadequacy of the HSM is its neglect of the inhomogeneous void-structure of the aerogel,
or more generally mesoscopic correlations that are observed in static structure factor, and to
which the superfluid transition temperature is sensitive.\cite{thu98,por99,sau03}
However, the transport properties are limited by the mean free path for
quasiparticles propagating ballistically within the aerogel, and hence are expected to be well
accounted for in the framework of the HSM since the geometric \emph{mfp} ($\ell$) is typically
much longer than the aerogel correlation length ($\xi_a$), e.g. $\ell/\xi_a\simeq 3$ for a
$98\%$ aerogel.
Possible corrections to transport processes resulting from a small distribution of large voids or to
fractal correlations on mesoscopic length scales $\lesssim\xi_a$ within the aerogel are not included
in the analysis presented in this work. However, the exact solution to the transport equation for
the two-channel scattering model discussed in this paper should be of value in identifying
observable corrections associated with correlated disorder or corrections to the two channel
scattering theory.

Liquid \He\ is a dense quantum liquid in which the interactions between Fermionic excitations
(quasiparticles) are one to two orders of magnitude larger than the mean kinetic energy per
particle.
These interactions lead to strongly renormalized branches of Fermionic excitations, reflecting
the correlated motion of many \He\ atoms, and the emergence of Bosonic excitations.
The Fermionic excitations bear resemblance to \He\ atoms only in terms of their quantum numbers for
spin ($s=\pm 1/2$) and fermion number ($e=\pm 1$).
The Bosonic excitations come with and without spin and can be understood in terms of pairs of
Fermionic excitations, e.g. the phonons of zero sound.
Finally, the coupling between the Bosonic and Fermionic excitations leads to finite lifetimes
for both types of excitations.\cite{lan56,lan57,lan59,abr58}

Interactions between \He\ quasiparticles enhance the collision rate for quasiparticles near the
Fermi surface, leading to a significant reduction in the lifetime of a quasiparticle at the
Fermi surface. However, Fermi statistics rescues the low-energy quasiparticles (as well as the
Bosonic modes).\footnote{This is not entirely true as relatively weak interactions between
quasiparticles with zero total momentum on the Fermi surface eventually leads to the Cooper
instability and to superfluidity at low temperatures, $k_B T_c \lll E_f$.} At low temperatures,
$T\ll E_f/k_B\approx 1\,\mbox{K}$, the number of excitations is low, $n_{\text{qp}}\approx (k_B
T/E_f) n$. Similarly, binary collision processes are confined to a small region of phase space
near the Fermi surface, $\Delta p\approx\,(k_B T/E_f)p_f$. As a result the Pauli exclusion
effect suppresses the quasiparticle collision rate,\cite{abr58}
\be\label{tau_in}
\frac{1}{\tau_{\text{in}}} = 
\frac{{m^{\star}}^3}{4\pi^4\hbar^6}\,\langle W \rangle\,\left(k_B T\right)^2 
\,\,\, \sim\,\,\, T^2 \,\,\,,
\ee
where $m^{\star}$ is the effective mass of a quasiparticle and $\langle W \rangle $ is the  
square of the transition matrix element for binary collisions averaged over the Fermi surface.

Furthermore, since the density of excitations is low, $n_{\text{qp}}\ll n$, transport
coefficients are given by formulae familiar from gas kinetic theory. In particular, the
transport of heat is dominated by thermally excited quasiparticles with the thermal conductivity
given by\cite{abr58}
\be\label{K}
\kappa = \frac{1}{3} (\bar{c}_v v_f)(v_f \tau_{\kappa})
\,,
\ee
where $\bar{c}_v=\frac{2\pi^2}{3}N_f k_B^2 T$ is the low-temperature specific heat, $N_f$ is the
quasiparticle density of states at the Fermi energy, $v_f$ is the Fermi velocity and
$v_f\tau_{\kappa}$ is the transport mean-free-path for heat conduction, with
$\tau_{\kappa}\sim\tau_{\text{in}}\sim T^{-2}$. Thus, heat transport becomes very
efficient in pure \He\ with $\kappa\sim 1/T$.\cite{abe67,gre84}

\begin{figure}[ht]
\epsfysize=0.85\hsize{\epsfbox{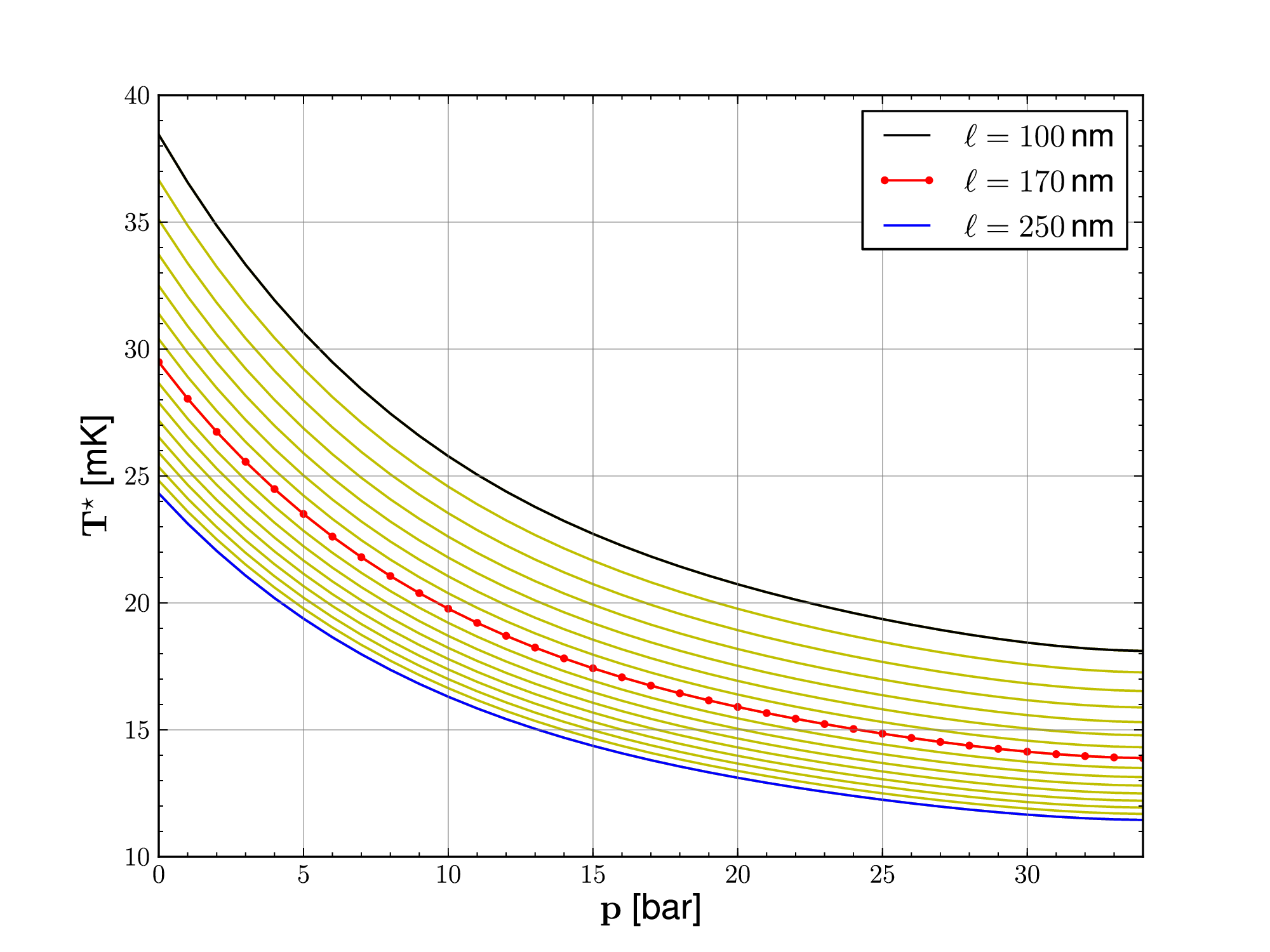}}
\caption{
Cross-over temperature vs. pressure for aerogels characterized by the \emph{mfp} as defined
by Eqs. \ref{tau_in}-\ref{Tstar}. The curves correspond to $\ell=100-250$ nm spaced by $10$ nm. The
red curve is for $\ell=170$ nm ($\varrho\approx 98\%$). The cross-over temperature is obtained from
Eq. \ref{Tstar} with $\langle\bar{W}\rangle$ given by Eqs. (\ref{W}-\ref{C_coefficients}), Table
\ref{table-CL_coefficients} (see the appendix in Sec. \ref{appendix-scattering_model}), and the
relevant Fermi-liquid parameters from Refs. \onlinecite{hal90,har00}.
}
\label{fig:Tstar}
\end{figure}

In aerogel, elastic collisions of \He\ quasiparticles with the silica strands lead to a
temperature-independent contribution to the mean free path and hence the quasiparticle
scattering rate. 
Thus, at sufficiently low temperatures, transport currents are limited by elastic scattering
from the aerogel, whereas inelastic scattering of quasiparticles dominates at high temperatures.
There is an intermediate regime where both mechanisms are important.
The cross-over temperature separating
these regimes is estimated from the inelastic collision rate in the pure \He\ in Eq.
(\ref{tau_in}) and the \emph{mfp} of the aerogel, $\ell$, which provides an estimate for the
elastic collision rate,
\be
\label{taul}
{\frac {1} {\tau_{\text{el}}}} = \frac {v_f}{\ell}\,\,=\,\,\mbox{constant}
\,.
\ee
Estimating $\tau_{\kappa}$ from 
Eqs. (\ref{tau_in}) and (\ref{taul}) gives,
\be
\kappa = \frac{1}{3}\bar{c}_v v_f^2 \tau_{\kappa}\,\sim\,
\Bigg\{ 
      \begin{matrix} 
       T & \quad ,\quad T < T_{\star}
       \cr
       1/T & \quad ,\quad T > T_{\star} 
       \,.
       \end{matrix}
\ee
The cross-over temperature, $T_{\star}$, defined by
$\tau_{\text{in}}(T_{\star})=\tau_{\text{el}}$, is given by\cite{rai98}
\be\label{Tstar}
T_{\star}=\frac{8E_f/k_B}{\sqrt{\pi k_f\ell\,\langle\bar{W}\rangle}}
\,,
\ee
where $\langle\bar{W}\rangle$ is the dimensionless quasiparticle transition probability averaged
over the Fermi surface (Eq. \ref{dimensionless_rate} of the appendix). For $98\%$ porosity aerogel
we estimate $\ell\approx 1700\,\mbox{\AA}$,\cite{thu98}, and thus $T_{\star}\approx 18$ mK at $p=15$
bar (see Fig. \ref{fig:Tstar}).
%

The cross-over from the high-temperature regime dominated by inelastic scattering to the
low-temperature regime dominated by elastic scattering off the aerogel is characteristic of all
transport processes for \He\ in aerogel.\cite{rai98,ven00,nom00,sau05} 
Below we report theoretical results for the heat transport coefficient of liquid
\Heaero, in the normal state, over the broad temperature range, $T_c \le T \ll
E_f/k_B$. Since the bulk properties of \He\ are well
known, measurements of the transport coefficients can provide quantitative information on the
effects of disorder on the transport of \He~quasiparticles through aerogel.
\section {Transport Theory}

In normal liquid \He~at low temperatures the transport of mass, energy and magnetization 
is carried predominantly by fermionic quasiparticles, whose distribution in phase space, 
$n_{\vp\sigma}$, is governed by the Boltzmann-Landau transport equation,\cite{abr57,bay91}
\be
\pder{n_{p}}{t}+\grad_{\vp}\varepsilon_{p}\cdot
                        \grad_{\vr}n_{p}
		       -\grad_{\vp}n_{p}\cdot
		        \grad_{\vr}\varepsilon_{p}
		       =\left(\pder{n_{p}}{t}\right)_{\text{coll}}
		       \equiv I_{p}
\,,
\ee
where $p=(\vp,\sigma)$ denotes the momentum and spin of the quasiparticles. For 
spin-independent transport the quasiparticle energy
\be
\varepsilon_{\vp}=\epsilon_{\vp}+\delta\varepsilon_{\vp}(\vr,t)+u_{\text{ext}}(\vp,\vr,t)
\,,
\ee
is the sum of the equilibrium excitation energy, $\epsilon_{\vp}$, the coupling to an external
scalar or vector potential, $u_{\text{ext}}(\vp,\vr,t)$, and the Landau molecular field energy.
The latter arises from the interaction of a quasiparticle with the distribution of
non-equilibrium quasiparticles,
\be
\delta\varepsilon_{\vp}=\sum_{\vp'\sigma'}\,f_{\vp,\vp'}\,\delta n_{\vp'}
\,,
\ee
where $\delta n_{\vp}$ is the deviation of the distribution function from (global) equilibrium,
\be
\delta n_{\vp}=n_{\vp} - n_0(\epsilon_{\vp})
\,.
\ee
For small disturbances from equilibrium the derivative of the equilibrium distribution function,
\be\label{derivative_Fermi}
-\pder{n_0}{\epsilon_{\vp}}=
\frac{1}{4k_B T}\sech^2\left(\frac{\epsilon_{\vp}-\mu}{2k_B T}\right)
\,,
\ee
confines the excitations to states that lie near the Fermi 
surface, $\epsilon_{\vp_f}=\mu$. The interaction energy between two quasiparticles is given 
by $f_{\vp,\vp'}$, and in contrast to the distribution function, varies slowly with $|\vp|$ in the 
vicinity of the Fermi surface. We can typically evaluate $f_{\vp,\vp'}$, as well as the density of 
states, $N(\epsilon)$, on the Fermi surface, i.e. $N(\epsilon)\simeq N_f$, and 
$2N_f f(p_f\hat{\vp},p_f\hat{\vp}')=F(\hat{\vp}\cdot\hat{\vp}')= 
\sum_{\ell}F_{\ell}\,P_{\ell}(\hat{\vp}\cdot\hat{\vp}')$. The latter equality defines the 
dimensionless Landau parameters. 
%
\subsection{Collision Integrals}

The right side of the transport equation, $I_{p}$, represents the change in the distribution
function resulting from collision processes. We consider two scattering processes for \He~in
aerogel: (i) elastic collisions of quasiparticles with ``impurities'' representing the aerogel
strands, and (ii) inelastic collisions between quasiparticles. The development of the transport
theory for \Heaero\ presented below, particularly the reduction of the transport equation in the low
temperature limit, parallels that development by Baym and Pethick in their review on transport in
pure liquid \He,\cite{bay78} and extends Brooker and Sykes' work on the transport coefficients
of Fermi liquids.\cite{syk70}

In our case the effects of aerogel scattering enter through a contribution to the collision
integral. For quasiparticle scattering by the aerogel strands
\be
\label{elcollinteg}
I^{\text{el}}_{p_1} = -\frac{2}{V}\,
{\sum_{p_2}}\,w(p_1,p_2)\,
\delta(\varepsilon_1-\varepsilon_2)\,[n_{p_1} - n_{p_2}]
\,,
\ee
where $w(p_1,p_2)$ is the transition rate for scattering of quasiparticles by the aerogel. 

For inelastic quasiparticle-quasiparticle collisions at low temperatures, $k_B T\ll E_f$, only
binary collisions are important. We denote ${\tt t}(p_1,p_2;p_3,p_4)$ as the scattering amplitude
for binary collisions between quasiparticles with momenta and spin $p_i=(\vp_i,\sigma_i)$. The
labels $p_1$ and $p_2$ refer to initial states while $p_3$ and $p_4$ refer to final states. Fermi's
Golden Rule for the transition rate $(p_1,p_2)\rightarrow(p_3,p_4)$ is:
\be
\Gamma = \frac{2\pi}{\hbar}\left|{\tt t}(p_1,p_2;p_3,p_4)\right|^2\,
         \delta(\varepsilon_1+\varepsilon_2-\varepsilon_3-\varepsilon_4)
\,.
\ee
For a translationally invariant system with spin-rotation invariant interactions between 
quasiparticles the transition rate includes momentum- and spin-conserving delta functions,
\ber
\Gamma =
	&\frac{1}{V^2}&W(p_1,p_2;p_3,p_4)
	\delta_{\vp_1+\vp_2,\vp_3+\vp_4}\,
	\delta_{\sigma_1+\sigma_2,\sigma_3+\sigma_4}
\nonumber\\
       &\times&
        \delta(\varepsilon_1+\varepsilon_2-\varepsilon_3-\varepsilon_4)
\,.
\eer
where $W$ is a smooth function of $\vp_i$.

The collision integral for binary scattering includes the phase space factors for collisions
that both increase and decrease the population of the state $p_1$ (scattering "in" and
scattering "out"). In particular,
\ber\label{inelcollinteg}
&&
I^{\text{in}}_{p_1} = 
-\sum_{p_2,p_3,p_4}
\Gamma(p_1,p_2;p_3,p_4)
\;\times
\\
&&
[n_{p_1} n_{p_2}(1-n_{p_3})(1-n_{p_4})
-
(1-n_{p_1})(1-n_{p_2})n_{p_3} n_{p_4}]
\nonumber
\,.
\eer
The sum over final states $(p_3,p_4)$ is restricted to avoid double counting of equivalent states 
of identical particles related by exchange of $p_3\leftrightarrow p_4$.

The collision integral vanishes when evaluated with a local equilibrium distribution function, 
i.e. $I[n^{l.e.}_{p}]\equiv 0$.
For the elastic scattering contribution to the collision integral (Eq. \ref{elcollinteg}) this identity is obvious. 
For the inelastic collision integral it is less so, but follows from the identity,
\ber\label{detailed_balance}
&&
\delta(\varepsilon_{p_1}+\varepsilon_{p_2}-\varepsilon_{p_3}-\varepsilon_{p_4})
\times
\\
&&
\left[
n_{0}(\varepsilon_{p_1})n_{0}(\varepsilon_{p_2})
(1-n_{0}(\varepsilon_{p_3}))(1-n_{0}(\varepsilon_{p_4}))
-
\right.
\nonumber \\
&&
\left.
n_{0}(\varepsilon_{p_3})n_{0}(\varepsilon_{p_4})
(1-n_{0}(\varepsilon_{p_1}))(1-n_{0}(\varepsilon_{p_2}))
\right]=0
\,,
\nonumber
\eer
where $n_0(\varepsilon)=1/(e^{\beta(\varepsilon-\mu)}+1)$ is the Fermi distribution. This identity
is a consequence of local equilibrium and the condition of detailed balance between the scattering
"in" and scattering "out" contributions to the collision rate.

Although translational invariance is violated by the presence of the aerogel medium,
the aerogel is sufficiently dilute that the scattering rate by the aerogel impurities is
typically small compared to excitation energies in the normal state, i.e. $\hbar/\tau_{\text{el}}\ll
k_B T$.\footnote{For 98\% aerogel this implies $T\gtrsim 1\,\mbox{mK}$.}
In this limit the effects of aerogel scattering on the intermediate states that enter
the inelastic collision integral can be neglected. Thus, momentum conservation holds for the
binary collision integral for normal \He~in high porosity aerogels. At lower temperatures, e.g.
in the superfluid phase,or for lower porosity aerogels, this approximation breaks down.
This limit requires a microscopic treatment of the effects of aerogel scattering on inelastic
collision processes which is outside the scope of the phenomenological Boltzmann-Landau
transport theory.
\subsection{Linearized Transport Equation}

The transport coefficients of liquid \He~in areogel are calculated from solutions of 
the {\sl linearized} transport equation in steady-state. The particular solution depends on the 
nonequilibrium conditions that are established. For small deviations from equilibrium the 
nonequilibrium steady state is specified by a {\sl local equilibrium} distribution function,
\be\label{local_equilibrium}
n^{l.e.}_{p}=
\frac{1}{1 + e^{(\varepsilon_{\vp}(\vr)-\mu(\vr))/k_BT(\vr)}}
\,,
\ee
parametrized by a local temperature, $T(\vr)$, chemical potential, $\mu(\vr)$ and  
quasiparticle energy, $\varepsilon_{\vp}(\vr)$. The transport equation naturally separates by 
expanding about this local equilibrium distribution,
\be
\delta \bar{n}_{p} = n_{p} - n^{l.e.}_{p}
\,,
\ee
because the collision integral vanishes under the conditions of local equilibrium, 
$I_{p}[n^{l.e.}_{p}]\equiv 0$. The linearized transport equation 
becomes,
\be\label{linear_transport1}
\vv_{\vp}\cdot\grad_{\vr}\,n_{p}^{l.e.} 
-
\left({\pder{n_{p}^{l.e.}}{\varepsilon_{\vp}}}\right)\vv_{\vp}\cdot\grad_{\vr}\,
                           \varepsilon_{\vp}=\delta I_{p}[\delta\bar{n}_p]
\,.
\ee
The left side of Eq. (\ref{linear_transport1}) supplies the driving terms, e.g. $\grad\mu$ 
and $\grad T$, for the collision terms on the right side that act to restore equilibrium. 
Using Eq. (\ref{local_equilibrium}) the linearized transport equation reduces to
\be\label{linear_transport}
-\left(\pder{n_0}{\varepsilon_{\vp}}\right)\, \vv_{\vp}\cdot
\Bigg[\left(\frac{\varepsilon_{\vp}-\mu}{T}\right)\grad T + \grad\mu\Bigg] =\delta
I_{p}[\delta\bar{n}_p]
\,,
\ee
where $\delta I_{p}$ is the collision integral to linear order in $\delta\bar{n}_{p}$.
\subsection{Quasiparticle Currents}

The mass and heat currents are determined by the solution for $\delta\bar{n}_{p}$ of 
Eq. (\ref{linear_transport}). In particular, the mass current is given by
\be\label{quasiparticle_mass_current}
\vj_{\mathsf m}=\sum_{\vp\sigma}\,{\mathsf m}^*\vv_{\vp}\,\delta \bar{n}_{\vp\sigma}
\,,
\ee
where ${\mathsf m}^*=p_f/v_f$ is the quasiparticle effective mass.
This form for the mass current is applicable to interacting Fermi liquids which are Galilean
invariant.\cite{bay91} In pure liquid $^3He$ quasiparticle-quasiparticle interactions which give
rise to the enhancement of the Fermionic mass are Galilean invariant. For liquid \Heaero\ Galilean
invariance is violated by quasiparticle scattering off the aerogel. However, the non-Galilean
contribution to the effective mass is of the order of concentration of scattering centers, $n_s/n
\ll 1$, and thus negligible compared to the quasiparticle-quasiparticle effective mass enhancement.

Similarly, the quasiparticle heat current is given by the transport of excitations 
with energy, $\xi_{\vp}=\varepsilon_{\vp}-\mu$,  
\be\label{quasiparticle_heat_current}
\vj_{\mathsf q}=\sum_{\vp\sigma}\,\xi_{\vp}\vv_{\vp}\,\delta \bar{n}_{\vp\sigma}
\,.
\ee
\subsection{Elastic Scattering Limit}

Transport properties of normal \He\ in aerogel at sufficiently low temperatures, i.e. $T \ll
T_{\star}$, are limited by elastic scattering of quasiparticles by the aerogel structure. In this
limit the transport equation is given by Eq. (\ref{linear_transport}) with the collision term of Eq.
(\ref{elcollinteg}). The integral equation for $\delta\bar{n}_{\vp}$ is
\ber\label{linear_transport-elastic}
-\left(\pder{n_0}{\xi_{\vp}}\right)\vv_{\vp}\cdot\vZ(\xi_{\vp}) = -\sum_{\vp'\sigma'}\,
&&
w(\vp,\vp')\,\delta(\xi_{\vp}-\xi_{\vp'})\,
\nonumber\\
\times
&&
\left[\delta\bar{n}_{\vp\sigma}-\delta\bar{n}_{\vp'\sigma'}\right]
\,,
\eer
where 
\be
\vZ(\xi_{\vp})=\left[\left(\frac{\xi_{\vp}}{T}\right)\grad T + \grad\mu\right]
\,.
\ee
The solutions to Eq. (\ref{linear_transport-elastic}) are determined by the energy and momentum
dependences of the driving term and are familiar from the theory of electron-impurity scattering in
metals.\cite{abrikosov88} To proceed further we need the \He\ quasiparticle-aerogel scattering
probability, $w(p,p')$.
\subsection{Elastic Scattering Model}

We model the aerogel as a distribution of local scattering centers represented 
by the potential, $U(\vr)=\sum_{i}u(\vr-\vR_i)$. The terms $u(\vr-\vR_i)$ represent the 
potential provided by the aerogel scattering centers at the {\sl fixed} positions, 
$\{\vR_i|i=1...N_s\}$. For a random distribution of uncorrelated scattering centers 
the rate is proportional to the mean number density of scattering centers, $n_s=N_s/V$. 
In the Born approximation the transition rate is related to the matrix elements of $u$,
\be
w(\vp\sigma;\vp'\sigma') =
n_s\frac{2\pi}{\hbar}\vert\bra{\vp'\sigma'}u\ket{\vp\sigma}\vert^2 
\,.
\ee
For stronger scattering the potential $u$ is replaced by the $t$-matrix for quasiparticle 
scattering by aerogel strands.\cite{sau03}

We shall assume that the scattering by the aerogel is {\it non-magnetic}. This should be sufficient
for describing transport processes in zero field, particularly if the aerogel strands are ``coated''
with a layer of solid \Hefour. However, it is known that \He~atoms form a highly polarizable solid
layer on the surface of the aerogel strands and that these nuclear spins exhibit a Curie-like spin
susceptibility.\cite{spr95} Thus, spin-exchange scattering of \He~quasiparticles by localized and
polarizable \He~spins may be relevant to magnetic transport processes and transport in relatively
low magnetic fields. 
%
%
The simplest scattering model for \Heaero\ assumes the \He\ quasiparticles interact with the 
aerogel via an isotropic scattering potential.\cite{thu98} There is no preferred direction 
within the aerogel and the scattering probability depends on the relative orientation of the 
initial and final quasiparticle momenta. In this case,
\be
w(\vp,\vp') =
w(\hat\vp\cdot\hat\vp';\xi_{\vp}) =
\sum_{l\ge 0}w_{l}(\xi_{\vp})\,\cP_{l}(\hat\vp\cdot\hat\vp')
\,,
\ee
where $w_{l}(\xi_{\vp})$ is the scattering probability for quasiparticles with relative orbital
angular momentum $l$, and $\cP_{l}(x)$ is the corresponding Legendre polynomial. Note that
$|\vp|=|\vp'|$ for elastic scattering, and we have parametrized the functional dependence of
$w(\vp,\vp')$ on $|\vp|$ by the energy $\xi_{\vp}=v_f(|\vp|-p_f)$ measured relative to the Fermi
surface. The probabilities for scattering in the orbital channels are proportional to,
\be
w_{l}(\xi_{\vp})\equiv\frac{1}{(2l+1)}\int_{-1}^{+1}\,
\frac{dx}{2}\,w(x;\xi_{\vp})\,\cP_{l}(x)
\,,
\ee
which vary smoothly with $\xi_{\vp}$ on the scale of $E_f$. 

For an isotropic scattering medium we make the {\sl ansatz},
\be\label{ansatz_elastic_scattering}
\delta\bar{n}_{\vp}=\left(\pder{n_0}{\xi_{\vp}}\right)
\left[\vv_{\vp}\tau_{\text{el}}(\xi_{\vp})\right]\cdot\vZ(\xi_{\vp})\,
\,.
\ee
The momentum sum is represented as
\be
\sum_{\vp'}(\ldots)=\int\dangle{\vp'}\,
                             \int d\xi_{\vp'}\,N(\xi_{\vp'})(\ldots)
\,.
\ee
The terms $(\partial n_0/\partial\xi_{\vp})$ and $\delta(\xi_{\vp}-\xi_{\vp'})$ 
confine $|\vp'|=|\vp|\simeq p_f+\xi_{\vp}/v_f$, so we obtain
\be
\frac{1}{\tau_{\text{el}}(\xi_{\vp})} = 2N(\xi_{\vp})\,
\langle
\,w(\hat\vp\cdot\hat\vp';\xi_{\vp})\left(1 - \hat\vp\cdot\hat\vp'\right)
\rangle_{\hat\vp'}
\,,
\ee
where $\langle\ldots\rangle_{\vp'}\equiv\int\frac{d\Omega_{\vp'}}{4\pi}(\ldots)$. 
We introduce the scattering rate for orbital channel $l$,
\be
\frac{1}{\tau_{l}(\xi_{\vp})} 
\equiv 
\frac{2N(\xi_{\vp})}{(2l+1)}\int_{-1}^{+1}\,\frac{dx}{2}\,w(x;\xi_{\vp})\,\cP_{l}(x)
=2N(\xi_{\vp})w_{l}(\xi_{\vp})
\,,
\ee
and express the transport scattering rate in terms of the $l=0$ and $l=1$ 
(s- and p-wave) scattering rates, 
\be
\frac{1}{\tau_{\text{el}}(\xi_{\vp})} 
= 
\frac{1}{\tau_{\text{0}}(\xi_{\vp})} 
- 
\frac{1}{\tau_{\text{1}}(\xi_{\vp})} 
\,.
\ee
Note that the density of states, $N(\xi_{\vp})$, and the scattering probabilities,  
$w_{l}(\xi_{\vp})$, vary slowly with excitation energy on the scale of the Fermi 
energy.
Thus, for many cases of interest we can safely neglect the energy dependence of 
the scattering rate and evaluate 
$\tau_{\text{el}}(\xi_{\vp})\simeq\tau_{\text{el}}(0)\equiv\tau_{\text{el}}$.

In particular, the mass current induced by a pressure gradient at constant temperature 
is determined by the quasiparticle {\sl mobility} defined by $\vj_{\text{m}} = -\nu\grad\mu$. 
For $k_B T\ll E_f$ Eq. (\ref{derivative_Fermi}) is 
sharply peaked at the Fermi energy, and the quasiparticle mobility calculated from 
Eqs. (\ref{quasiparticle_mass_current}) and (\ref{ansatz_elastic_scattering}) is to leading 
order in $T/E_f$,
\be
\nu=\twothirds N_f\,p_f\, (v_f\tau_{\text{el}})
\,,
\ee
which has the intuitive interpretation of transport of momentum $p_f$ over a distance of 
order the {\sl mfp}, $\ell=v_f\tau_{\text{el}}$, within the aerogel.
Similarly, the heat current, Eq. (\ref{quasiparticle_heat_current}), induced by an temperature 
gradient, $\vj_{\text{q}}=-\kappa\grad T$, defines the thermal conductivity, which to leading 
order in $T/E_f$ is given by,
\be\label{thermal_conductivity_elastic}
\kappa_{\text{el}}=\frac{2\pi^2}{9}\,k_B\,N_f\,(v_f k_B T)\,\ell
\,,
\ee
and also has the simple interpretation as the flux of thermal  energy $v_f k_BT$ transported 
over a distance of order $\ell$.

\subsection{Two Channel Collision Integral} 

For higher temperatures, i.e. $T\gtrsim T_{\star}$, both elastic and inelastic collision 
processes limit transport currents. The transport coefficients are then calculated from 
\be\label{linear_transport_equation}
-\left(\pder{n_0}{\xi_{p_1}}\right)\,
\vv_{p_1}\cdot
\Bigg[\left(\frac{\xi_{p_1}}{T}\right)\grad T + \grad\mu\Bigg]
=\delta I^{\text{in}}_{p_1}+\delta I^{\text{el}}_{p_1}
\,,
\ee
where the right side of Eq. (\ref{linear_transport_equation}) contains the linearized 
collision integrals for both inelastic and elastic scattering.
%
%
The elastic collision integral follows immediately from Eq. (\ref{elcollinteg}),
\be\label{elastic_collision_integral}
\delta I^{\text{el}}_{p_1} = 
-\sum_{p_2}\,w(p_1,p_2)\,
\left[\delta\bar{n}_{p_1} - \delta\bar{n}_{p_2}\right]
\delta(\xi_{p_1}-\xi_{p_2})
\,.
\ee
Since the driving terms in Eq. (\ref{linear_transport_equation}) are confined  to 
excitation energies within $k_B T$ of the Fermi energy we express
\be\label{1}
\delta\bar{n}_{p_i}=\left(\pder{n_0}{\xi_{p_i}}\right)\,\psi_{p_i}
\,,
\ee
where $\psi_{p_i}$ measures the deviation of the equlibrium distribution for 
quasiparticles with excitation energy $\xi_{p_i}$ at the point $\hat\vp_i$ on 
the Fermi surface. 

For $|\xi_{\vp_1}|\ll E_f$ energy conservation, combined with the phase-space 
restriction required by the Fermi distribution factors in Eq. (\ref{elastic_collision_integral}), 
forces the scattered excitation energy to be confined to the low-energy shell, 
i.e. $|\xi_{\vp_2}|\lesssim k_B T$. 
Thus, slowly varying functions of $\vp_i$ can be be evaluated with momenta, 
$|\vp_1|=|\vp_2|\simeq p_f + \xi_{\vp_1}/v_f$, i.e. in close vicinity of the Fermi 
momentum. The scattering rate $w$ reduces to a function of the {\sl directions} 
of the momenta for incident and scattered excitations with excitation energies 
near the Fermi energy, $w(\vp_1,\vp_2) \leadsto w(\hat\vp_1,\hat\vp_2;\xi_{\vp_1})$. 
For the isotropic scattering model and a driving term proportional to 
$\vv_{\vp_i}\cdot\grad T$ we set 
\be
\psi_{\vp_i} = \vv_{\vp_i}\cdot\grad T\;\varphi(\xi_{\vp_i})
\,, 
\ee
in which case the elastic collision integral becomes,
\be\label{elastic_collision_integral2}
\delta I^{\text{el}}_{p_1} = \frac{\vv_{\vp_1}\cdot\grad T}{T}\,
n_0(\xi_{\vp_1})(1-n_0(\xi_{\vp_1}))\,
\frac{1}{\tau_{\text{el}}(\xi_{\vp_1})}\;\varphi(\xi_{\vp_1})
\,.
\ee


Linearizing the inelastic collision rate in Eq. (\ref{inelcollinteg}), and
making use of Eq. (\ref{detailed_balance}) yields,
\be
\begin{split}
\label{incoll_linear}
\delta I^{\text{in}}_{p_1} = 
			&\frac{1}{T}\sum_{p_2,p_3,p_4}
			W(p_1,p_2;p_3,p_4)\,
	                \delta_{\sum_i\vp_i}
			\delta_{\sum_i\sigma_i} 
			\delta_{\sum_i\xi_{p_i}} 
\\	
			\times
			&\left[
			n_0(\xi_{p_1}) 
			n_0(\xi_{p_2}) 
			(1-n_0(\xi_{p_3})) 
			(1-n_0(\xi_{p_4}))
			\right]
\\	
			&\left[
			\psi_{p_1} + 
			\psi_{p_2} - 
			\psi_{p_3} - 
			\psi_{p_4}
			\right]
\,.
\end{split}
\ee

Here we consider an un-polarized Fermi liquid in which the only spin-dependent 
interactions are those that arise from exchange symmetry. In this case, 
$\xi_{p_i}=\xi_{\vp_i}$ and the distribution functions are independent of $\sigma_i$. 
Thus, we can carry out the sum over the spin states $\sigma_{2,3,4}$. 
We can also eliminate one of the momentum sums, resulting in,
\be\label{incoll_linear_spin_sum}
\begin{split}
\delta I^{\text{in}}_{p_1}=
			&\frac{2}{T}
			\sum_{\vp_2,\vp_3}
			W(\vp_1,\vp_2;\vp_3,\vp_4)
			\delta(\xi_{\vp_1}+\xi_{\vp_2}-\xi_{\vp_3}-\xi_{\vp_4})
\\
			\times
			&\left[
			n_0(\xi_{\vp_1}) 
			n_0(\xi_{\vp_2}) 
			(1-n_0(\xi_{\vp_3})) 
			(1-n_0(\xi_{\vp_4}))
			\right]
\\
			\times
			&\left[
			\psi_{\vp_1} + 
			\psi_{\vp_2} - 
			\psi_{\vp_3} - 
			\psi_{\vp_4}
			\right]
\,,
\end{split}
\ee
where $\vp_4=\vp_1+\vp_2-\vp_3$ and 
\be
\label{spinsum}
W=\onefourth W_{\uparrow\uparrow} + \onehalf W_{\uparrow\downarrow}
\,,
\ee
is the spin-averaged scattering rate; $W_{\uparrow\uparrow}$ is the scattering rate 
for $\sigma_1=\sigma_2=\uparrow$ and $W_{\uparrow\downarrow}$ is the rate for 
$\sigma_1=-\sigma_2=\uparrow$. The weight factors take into account the restriction to 
avoid double counting of equivalent states, so the remaining momentum sums over $\vp_3$ 
and $\vp_4$ are unrestricted.\cite{bay91}

For $|\xi_{\vp_1}|\ll E_f$ the energy and momentum conservation laws, combined 
with the phase-space restrictions required by the Fermi distribution factors in 
Eq. (\ref{incoll_linear_spin_sum}), force all excitation energies to be confined to the 
low-energy shell, i.e. $|\xi_{\vp_i}|\lesssim k_B T$. In this
limit slowly varying functions of $\vp_i$ can be evaluated on the Fermi surface. 
Thus, the scattering rate $W$ becomes a function of the {\sl directions} of the 
momenta for quasiparticles on the Fermi surface,
\be
W(\vp_1,\vp_2;\vp_3,\vp_4) \leadsto W(\hat\vp_1,\hat\vp_2;\hat\vp_3,\hat\vp_4)
\,,
\ee
and the momentum sums can be approximated by
\be
\sum_{\vp_i}(\ldots)\leadsto N_f\int\dangle{\vp_i}\,\int d\xi_{\vp_i}(\ldots)
\,.
\ee

To carry out the angular integrations we adopt Abrikosov and Khalatnikov's 
parametrization\cite{abr57} of $W$ in terms of the angle $\theta$ between the two incoming momenta, 
and $\phi$, the scattering angle between the planes defined by  
$\vn=\hat\vp_1\times\hat\vp_2$ and $\vn'=\hat\vp_3\times\hat\vp_4$.
The integration over the direction $\hat\vp_3$ is expressed in terms of angles 
relative to the conserved direction of the total momentum, $\vP=\vp_1+\vp_2$,
\be
\int\dangle{\vp_3}(\ldots) = 
\int_0^{2\pi}\frac{d\phi_3}{2\pi}\int_{-1}^{+1}\frac{d(\cos\alpha)}{2}(\ldots)
\,.
\ee
Since $p_4=\sqrt{P^2+p_3^2-2\vP\cdot\vp_3}$ we have 
$dp_4=(p_3/p_4)P\,d(\cos\alpha)$, and for momenta near the Fermi surface,
 $P\simeq 2p_f\cos(\theta/2)$. Thus, 
\be
d(\cos\alpha)\simeq\frac{d\xi_{\vp_4}}{2v_fp_f\cos(\theta/2)} 
\,.
\ee
Also, the azimuthal angle $\phi_3$ is the scattering angle up to a fixed 
but arbitrary constant, thus, $d\phi_3=d\phi$.
In the case of the integration over the incoming momentum direction 
$\hat\vp_2$ we choose the remaining momentum direction $\vp_1$ as the polar axis,
\be
\int\dangle{\vp_2}(\ldots)=
\int_{0}^{2\pi}\frac{d\phi_2}{2\pi}
\int_{-1}^{+1}\frac{d(\cos\theta)}{2}
(\ldots)
\,.
\ee

The binary collision integral then reduces to
\begin{widetext}
\be\label{inelcollint2}
\begin{split}
\delta I^{\text{in}}_{p_1}=
			&\frac{2}{T}\,N_f^2\left(\frac{1}{2v_fp_f}\right)
			\int d\xi_{\vp_2}
			\int d\xi_{\vp_3}
			\int d\xi_{\vp_4}
			\delta(\xi_{\vp_1}+\xi_{\vp_2}-\xi_{\vp_3}-\xi_{\vp_4})
			\left[
			n_0(\xi_{\vp_1}) 
			n_0(\xi_{\vp_2}) 
			(1-n_0(\xi_{\vp_3})) 
			(1-n_0(\xi_{\vp_4}))
			\right]
\\
			\times
			&
			\int_{-1}^{+1}\frac{d(\cos\theta)}{2}
			\int_{0}^{2\pi}\frac{d\phi}{2\pi}
			\left(\frac{W(\theta,\phi)}{2\cos(\theta/2)}\right)
			\int_{0}^{2\pi}\frac{d\phi_2}{2\pi}
			\left[
			\psi_{\vp_1} + 
			\psi_{\vp_2} - 
			\psi_{\vp_3} - 
			\psi_{\vp_4}
			\right]
\,.
\end{split}
\ee
\end{widetext}

The inhomogeneous terms of the linearized transport equation dictate the symmetry of the 
solution with respect to excitation energy, $\xi_{\vp_1}\rightarrow -\xi_{\vp_1}$, and momentum 
direction, $\hat\vp_1$.
We separate the angular and energy dependences of the non-equilibrium distribution 
function with the {\sl ansatz}, $\psi_{\vp_i}=\psi^{(+)}_{\vp_i}+\psi^{(-)}_{\vp_i}$,
\ber
\psi^{(+)}_{\vp_i}	&=& \vv_{\vp_i}\cdot\grad \mu\;\varphi^{(+)}(\xi_{\vp_i})
\\
\psi^{(-)}_{\vp_i}	&=& \vv_{\vp_i}\cdot\grad T\;\varphi^{(-)}(\xi_{\vp_i})
\,,
\eer
where $\varphi^{(\pm)}(\xi_{\vp_i}) =\pm\varphi^{(\pm)}(-\xi_{\vp_i})$. We can now carry 
out the integration over $\phi_2$ for each term in  Eq. (\ref{inelcollint2}),
\be
\hspace*{-3mm}
\int_0^{2\pi}\frac{d\phi_2}{2\pi}\psi_{\vp_i} =  
x_i\vv_{\vp_1}\cdot
\left[
\grad\mu 
\varphi^{(+)}(\xi_{\vp_i})
+
\grad T
\varphi^{(-)}(\xi_{\vp_i})
\right],
\ee
where the direction cosines, $x_i=\hat\vp_i\cdot\vp_1$ for $i=2,3,4$, are simply related 
to $(\theta,\phi)$. The angular integrations 
decouple from the energy integrations which are confined to the low-energy shell 
near the Fermi surface. We define the average
scattering rate
\be\label{Fermi-surface_average}
\langle W \rangle \equiv 
			\int_{-1}^{+1}\frac{d(\cos\theta)}{2}
			\int_{0}^{2\pi}\frac{d\phi}{2\pi}
			\left(\frac{W(\theta,\phi)}{2\cos(\theta/2)}\right)
\,,
\ee
as well as the weighted averages,
$r_i = \langle x_i W \rangle/\langle W \rangle$.
Changing $\xi_{\vp_3,\vp_4}\rightarrow - \xi_{\vp_3,\vp_4}$ the resulting collision 
integral reduces to 
\begin{widetext}
\be
\label{inelcollint6}
\delta I^{\text{in}(\pm)}_{p_1}
			= \frac{1}{T}\,\frac{N_f^2}{p_f v_f}\langle W \rangle\vv_{\vp_1}\cdot
			\begin{pmatrix}
			  \grad\mu \cr \grad T
			\end{pmatrix}
			\times
			n_0(\xi_{\vp_1})
			\times
			\Bigg[
			I(\xi_{\vp_1})\,\varphi^{(\pm)}(\xi_{\vp_1})\,
			+\lambda^{(\pm)}
			\int d\xi_{\vp_2}
			n_0(\xi_{\vp_2})
			K(\xi_{\vp_1}+\xi_{\vp_2})\,\varphi^{(\pm)}(\xi_{\vp_2})\,
			\Bigg]
\,,
\ee
\end{widetext}
where
\ber
\label{integral_K}
K(\xi)      &=&\int d\xi_3\int d\xi_4\,\delta(\xi+\xi_3+\xi_4)\,n_0(\xi_3) n_0(\xi_4)
\nonumber\\
            &=& \left(\frac{\xi}{1-e^{-\xi/T}}\right) \,,
\\
\label{integral_I}
I(\xi)       &=&\int d\xi_2\,n_0(\xi_2) K(\xi+\xi_2)
\nonumber\\
            &=& \onehalf\left(\pi^2 T^2+\xi^2\right)\left(1-n_0(\xi)\right)
\,,
\eer
and $\lambda^{(\pm)}$ are given by,
\be
\lambda^{(\pm)} \equiv r_2\mp(r_3+r_4) 
= \Bigg\{
  \begin{matrix} 
   -1 \cr \langle W (1+2\cos\theta)\rangle/\langle W \rangle
   \,.
  \end{matrix}
\ee
It is then convenient to measure the excitation energy in units of $T$, 
i.e. set $t_i\equiv\xi_{\vp_i}/T$, for $i=1,2,3,4$. The resulting inelastic 
collision integrals reduce to
\begin{widetext}
\be
\label{inelcollint7}
\delta I^{\text{in}(\pm)}_{p_1}
			=
			\frac{1}{T}\,
			n_0(t_1)
			\vv_{\vp_1}\cdot
			\begin{pmatrix}
			  \grad\mu \cr \grad T
			\end{pmatrix}
			\frac{1}{\tau_{\text{in}}}
			\times
			\Bigg[
			\onehalf
			(1 - n_0(t_1))\,
			\left(\pi^2+t_1^2\right)
			\varphi^{(\pm)}(t_1)\,
			+\lambda^{(\pm)}
			\int dt_2\,
			n_0(t_2)\,
			\frac{t_1+t_2}{\sinh\left(\frac{t_1+t_2}{2}\right)}\,
			\varphi^{(\pm)}(t_2)\,
			\Bigg]
\,,
\ee
\end{widetext}
where $1/\tau_{\text{in}}$ is the quasiparticle-quasiparticle collision rate given 
in Eq. (\ref{tau_in}). These same transformations applied to the elastic collision 
integral in Eq. (\ref{elastic_collision_integral2}) yield, 
\be\label{Iel}
\delta I^{\text{el}(\pm)}_{p_1}=
			\frac{1}{T}\,n_0(t_1)(1-n_0(t_1))
			\vv_{\vp_1}\cdot
			\begin{pmatrix}
			  \grad\mu \cr \grad T
			\end{pmatrix}
			\frac{1}{\tau_{\text{el}}}
			\times
			\varphi^{(\pm)}(t_1)
\,,
\ee
where $1/\tau_{\text{el}}$ is the rate for quasiparticles on 
the Fermi surface scattering elastically off the aerogel.
Finally, the left-hand side of the linearized transport equation
provides the driving terms that determine the particular solution
for the nonequilbrium distribution functions. The driving terms which 
are even and odd under $t_1\rightarrow -t_1$ are 
\be
L^{(\pm)}(t_1) = \frac{1}{T}\,n_0(t_1)\left(1-n_0(t_1)\right)\,
		\vv_{\vp_1}\cdot
		\begin{pmatrix}
		  \,\:\:\grad\mu \cr t_1 \grad T
		\end{pmatrix}
\,.
\ee
We simplify the even and odd components of the transport equation by
an additional transformation of the distribution function,
\be
\zeta^{(\pm)}(t)
\equiv 
\frac{1}{\tau_{\text{in}}}\,
\frac{\varphi^{(\pm)}(t)}{2\cosh(t/2)}\,
\,.
\ee
The linearized transport equation with both inelastic and elastic channels 
included in the collision integral then reduces to the linear integral equations,\cite{smith89}
\begin{multline}
\label{two-channel_transport_equation}
	\frac{1}{\cosh(t_1/2)}
	\begin{pmatrix} 1 \cr t_1 \end{pmatrix}
	= 
	\left(t_1^2+\pi^2+\frac{2\tau_{\text{in}}}{\tau_{\text{el}}}\right)\zeta^{(\pm)}(t_1)
\\
	\mp\lambda^{(\pm)}
	\int dt_2\,\left(\frac{t_1-t_2}{\sinh((t_1-t_2)/2)}\right)
	\zeta^{(\pm)}(t_2)
\,,
\end{multline}

Physical solutions for $\zeta^{(\pm)}$ are non-vanishing only in the low-energy region
near the Fermi level, i.e. $|t| \ll 1$. Thus, we can Fourier transform,
\be
{\tilde\zeta}^{(\pm)}(z)\equiv\int_{-\infty}^{+\infty}\,dt e^{-izt}\,\zeta^{(\pm)}(t)
\,,
\ee
and convert the integral equation for $\zeta^{(\pm)}(t)$ into a linear differential 
equation for $\tilde{\zeta}^{(\pm)}(z)$,
\be
\begin{split}\label{transport_differential_equation}
	\left(
	\pder{^2}{z^2} 
	- 
	\pi^2\gamma^2 
	\right)
	&\tilde{\zeta}^{(\pm)}(z)
	\mp 
	2\pi^2\lambda^{(\pm)}\sech^2(\pi z)\,\tilde{\zeta}^{(\pm)}(z)
	\\
	& = -2\pi\,\sech(\pi z) 
	\begin{pmatrix} 
	  1
	  \cr
	  -i\pi\tanh(\pi z)
	\end{pmatrix}
	\,,
\end{split}
\ee
where 
\be\label{structure_constant}
\gamma\equiv\sqrt{1+\frac{2}{\pi^2}\frac{\tau_{\text{in}}}{\tau_{\text{el}}}}
\,.
\ee
We cast this differential equation into standard form 
defined on the domain $x\in [-1,1]$ with the transformation,
\ber
\tanh(\pi z) &=& x \,,
\\
\tilde\zeta^{(\pm)}(z) &=& \cZ^{(\pm)}(x) 
\,,
\eer
and the differential operator,
\be
\cD[f]\equiv \der{}{x}\left((1-x^2)\der{f}{x}\right) 
- \frac{\gamma^2}{(1-x^2)}\,f
\,.
\ee
Thus, the nonequilibrium distribution function is obtained as the solution of 
an inhomogeneous linear differential equation,
\be\label{differential_transport}
\begin{split}
\cD[\cZ^{(\pm)}] 
\mp 
2\lambda^{(\pm)}\cZ^{(\pm)} 
=\,R^{(\pm)}(x) 
\,,
\end{split}
\ee
where
\be\label{inhomogeneous_terms}
R^{(\pm)} = \frac{1}{\sqrt{1-x^2}}
\begin{pmatrix}
	-2/\pi 
	\cr
	+2i\,x
\end{pmatrix}
\,.
\ee

\subsection{Thermal Conductivity}

The heat current for example can be expressed in terms of a particular 
solution of Eq. (\ref{differential_transport}),
\be
\begin{split}
	\vj_{\text{\sf q}}
&
\simeq	
\,2N_f \int\dangle{\vp}\int d\xi_{\vp}\left(v_f\hat\vp\,\xi_{\vp}/T\right)
\\
&
\times
n_0(\xi_{\vp})
\left(1 - n_0(\xi_{\vp})\right)\,
\psi_{\hat\vp}(\xi_{\vp})
\,.
\end{split}
\ee
Carrying out the transformation from $\psi_{\hat\vp}(\xi_{\vp})\rightarrow \cZ(x)$ 
we obtain the following expression for the thermal conductivity,
\be
\kappa = \onethird (\bar{c}_{\text{V}}v_f)(v_f\tau_{\text{in}})\,
         S_{\text{$\kappa$}}(T)
\,,
\ee
where
\be
S_{\text{$\kappa$}}(T) \equiv
\frac{3}{4\pi^2}\int_{-1}^{+1}\,dx 
\,R^{(-)}(x)\,\cZ^{(-)}(x)
\,.
\ee

The particular solution for $\cZ^{(-)}$ is obtained as an expansion,
\be
\cZ^{(-)} = \sum_{n}\,\cA_{n}\phi_{n}(x)
\,,
\ee
in the complete set of orthonormal eigenfunctions, $\{\phi_{n}(x)\}$, of the 
homogeneous differential equation,
\be\label{eigenvalue_equation}
\cD[\phi_{n}] + 2 \alpha_{n}\phi_{n} = 0
\,,
\ee
where $\alpha_n$ is the eigenvalue associated with the eigenfunction, $\phi_n(x)$.
The coefficients, $\{\cA_n\}$, are determined from Eqs. (\ref{differential_transport}), 
(\ref{inhomogeneous_terms}) and (\ref{eigenvalue_equation}) and the 
orthogonality condition,
\be\label{orthonormality}
\int_{-1}^{+1}dx\,\phi_{n'}^{*}(x)\phi_{n}(x)
=
\delta_{n'n}
\,.
\ee
In particular,
\be
\cA_{n} = \frac{c_{n}}{\lambda_{\kappa}-\alpha_{n}}
\,.
\ee
where
\be
c_{n} = \int_{-1}^{+1}dx\,\phi_{n}^{*}(x)R^{(-)}(x)
\,,
\ee
is the overlap of the $n^{\text{th}}$ eigenfunction with the driving term in the 
transport equation. Also, we set $\lambda^{(-)}\equiv\lambda_{\kappa}$ 
above and hereafter.
The same term appears in the kernel of the heat current. Thus, the thermal 
conductivity, in particular, $S_{\text{$\kappa$}}(T)$, is determined by the weighted 
sum over the eigenvalue spectrum,
\be\label{sum_kappa_spectrum}
S_{\text{$\kappa$}}(T) =
\frac{3}{8\pi^2}\sum_{n}
\frac{|c_n|^2}{\alpha_n - \lambda_{\kappa}}
\,,
\ee

\subsection{Pure Fermi-Liquid}\label{Pure_Fermi-Liquid}

For pure \He\ ($\tau_{\text{el}}\rightarrow\infty$) Eq. (\ref{eigenvalue_equation}) 
reduces to the differential equation for the associated Legendre functions,\cite{syk70,smith89}
\be
(1-x^2)\der{^2\phi_n}{x^2} - 2x\der{\phi_n}{x}
\left(2\alpha_n- \frac{1}{(1-x^2)}\right)\,\phi_n = 0
\,.
\ee
The bounded, odd-parity solutions, relevant to heat transport, on the 
domain $x\in[-1,+1]$, are the associated Legendre polynomials, 
$P_n^{1}(x)$ with eigenvalues, $2\alpha_n=n(n+1)$ for $n=1,2,3,\ldots$. 
The standard orthognality relation for the polynomials is\cite{abramowitz72}
\be
\hspace*{-3mm}
\int_{-1}^{+1}dx\,P^1_{n}(x)^{*}\,P^1_{n'}(x) = \frac{2n(n+1)}{2n+1}\,\delta_{nn'}
\,,\quad n\ge 1
\,.
\ee
The evaluation of the overlap integrals leads to the solution for the thermal 
conductivity (Eq. (\ref{K})) obtained by Brooker and Sykes,\cite{bro68,syk70} 
and independently by Jensen et al.\cite{jen68}, with transport time 
$\tau_{\text{$\kappa$}} = \tau_{\text{in}}\,S^{\text{$\infty$}}_{\text{$\kappa$}}$, 
where
\be\label{Sum_kappa_pure}
\hspace*{-3mm}
S^{\text{$\infty$}}_{\text{$\kappa$}} = 
	\frac{6}{\pi^2}\sum_{m\ge 2}^{\text{even}}
	\left(\frac{2m+1}{[m(m+1)-2\lambda_{\kappa}]m(m+1)}\right)
\,,
\ee
which depends on the angular average of the scattering amplitude via $\lambda_{\text{$\kappa$}}$ and
is independent of temperature. Thus, the thermal conductivity diverges as $\kappa\propto 1/T$ as
$T\rightarrow 0$ because the number of thermal excitations, the heat capacity and the number of
final states for binary collisions are all vanishing as $T$. 
Note that $\lambda_{\kappa}=\langle W(1+2\cos\theta)\rangle/\langle W\rangle$ is a measure of the
relative importance of forward vs. backscattering, and is restricted to the domain, $-1 <
\lambda_{\kappa} < 3$. The resulting spectral sum, $S^{\text{$\infty$}}_{\text{$\kappa$}}$, is
finite since $\lambda_{\kappa} < 3$. However, at any fixed temperature $\kappa$ increases
dramatically for quasiparticle scattering that is predominantly in the forward direction, i.e. for
$\lambda_{\kappa}\rightarrow 3$. Note also that
$S^{\text{$\infty$}}_{\text{$\kappa$}}\rightarrow
\frac{5}{6\pi^2}\frac{1}{1-\tinyonethird\lambda_{\kappa}}$ in this limit.

\subsection{Exact Solution}\label{sec:exact_solution}

Here we extend the exact solution for pure \He\cite{bro68,jen68} to that of \He\ in aerogel
described by the two-channel collision integral for binary quasiparticle collisions and
quasiparticle-aerogel scattering. 
Bennett and Rice extended the analysis of Refs. \onlinecite{bro68,jen68} to collisional scattering
of s- and d-electrons combined with electron-impurity scattering.\cite{ben69} Their results for the
electrical and thermal conductivity are expressed as a sum over weighted integrals of products of
Gegenbauer polynomials. 
Our analysis also starts from a two-channel extension of the integral equation of Refs.
\onlinecite{bro68,jen68}, i.e. Eq. (\ref{two-channel_transport_equation}). The results presented
below are a closed form analytic solution to the linearized Boltzmann equation and thermal
conductivity, and an exact perturbation theory result for the inelastic corrections to the elastic
limit which is used to develop a very accurate approximate solution for the thermal conductivity
that is valid for all temperatures (above $T_c$), pressures and aerogel densities, \underline{and}
is fast and easy to evaluate.

Elastic scattering by the aerogel modifies the form of the
eigenfunctions for the nonequilibrium distribution function, and leads to an eigenvalue spectrum
that varies strongly with temperature. The key parameter is the structure constant ($\gamma$) in
Eqs. (\ref{transport_differential_equation}) and (\ref{structure_constant}) in the differential
equation for the distribution function. The temperature dependence is conveniently exhibited by
scaling Eq. (\ref{structure_constant}) in terms of the cross-over temperature, $T_{\star}$, defined
in Eq. (\ref{Tstar}),
\be\label{structure_constant_Tstar}
\gamma = \sqrt{1+\frac{2}{\pi^2}\left(\frac{T_{\star}}{T}\right)^2}
\,.
\ee

The eigenfunctions of Eq. (\ref{eigenvalue_equation}) for any $\gamma >1$ 
must be bounded on the interval $[-1,+1]$. The singular points at $x=\pm 1$ 
have indicial equations with one physically allowed solution in the 
neighborhood of the singular point; in particular, since $\gamma >1$ 
we select the physical solutions which must behave as
\ber
\phi_n &\sim& (1-x)^{\gamma/2}\,,\quad x\sim +1
\\
\phi_n &\sim& (1+x)^{\gamma/2}\,,\quad x\sim -1
\,.
\eer
Thus, we extract the behavior near the singular points and express
\be
\phi_n(x) = (1-x^2)^{\gamma/2}\,g_n(x)
\,,
\ee
where $g_n(x)$ is analytic on the domain $[-1,+1]$, and governed 
by the differential equation,
\be
(1-x^2)\der{^2 g_n}{x^2} - 2(\gamma+1)x\der{g_n}{x} + 
[2\alpha_n-\gamma(\gamma+1)]g_n =0
\,.
\ee
Analytic solutions on the finite domain can be represented as a 
Taylor expansion about $x=0$,
\be
g_n(x)=\sum_{m=0}^{\infty}\,G_m\,x^m
\,.
\ee
The differential equation determines the recurrence formula for the coefficients,
\ber
&&G_{m+2}=\frac{N_m}{(m+2)(m+1)}\,G_m\,,
\\
&&N_m=\left[m(m-1)+2m(\gamma+1)+\{\gamma(\gamma+1)-2\alpha_m\}\right]
\nonumber\,.
\eer
Thus, the solutions break up into even and odd parity solutions depending 
on the coefficients $G_0$ and $G_1$. For even parity solutions, we set 
$G_0\ne 0$ and generate the solutions from the recurrence relation:
\be
G_m=\frac{ N_0 N_2 \ldots N_{m-2}}{m!}\,G_0\,,\quad m=2,4,\ldots
\,.
\ee
Similarly, for the odd-parity solutions we start from $G_1\ne 0$ and find
\be
G_m=\frac{ N_1 N_3 \ldots N_{m}}{(m+2)!}\,G_1\,,\quad m=3,5,\ldots
\,.
\ee

In either case
\be
\lim_{m\rightarrow\infty}\frac{G_{m+2}}{G_m}\rightarrow 1
\,.
\ee
Thus, the series solution diverges at $|x|=1$ {\sl unless} the expansion 
truncates at a finite value of $m$. This restricts the physical solutions for 
$g_n(x)$ 
to a set of polynomials, and an eigenvalue spectrum determined by the condition: 
\be
G_{n+2} = 0\leadsto N_{n}=0
\,.
\ee
Expressing the eigenvalue as $2\alpha_n=\epsilon_n(\epsilon_n+1)$, we obtain
\be
\epsilon_n=\gamma + n \,,\quad n=0,2,4,\ldots\,(n=1,3,5,\ldots)
\,,
\ee
for even (odd) parity solutions. The corresponding eigenfunctions are
\be
\phi_n(x)=(1-x^2)^{\gamma/2}\,\sum_{m}^{n}G_m^{(n)}x^m
\,,
\ee
with the summation over even (odd) integers for even (odd) parity 
eigenfunctions. The coefficients can be expressed in terms of Gamma 
functions. In particular, for the odd-parity eigenfunctions, which are relevant for computing the thermal conductivity,
\begin{multline}
C_m^{(n)}\equiv G_m^{(n)}/G_1^{(n)} = (-1)^{(m-1)/2}\,\frac{2^{m-1}}{m!}
\\
\times
\frac{\Gamma(\frac{n}{2}+\onehalf)}{\Gamma(\frac{n-m}{2}+1)}
\frac{\Gamma(\gamma+\frac{n}{2}+1+\frac{m-1}{2})}{\Gamma(\gamma+\frac{n}{2}+1)}
\,.
\end{multline}
Coefficient $G_1^{(n)}$ is fixed by the normalization of $\phi_n(x)$, 
\be
\left|G_1^{(n)}\right|^2 = \left[\sum_{m=1}^{n}\sum_{p=1}^{n} 
C_m^{(n)} C_p^{(n)} \cB(\gamma+1,\frac{m+p+1}{2})
\right]^{-1}
\,,
\ee
and $\cB(x,y)=\Gamma(x)\Gamma(y)/\Gamma(x+y)$ is the 
Beta function.\cite{abramowitz72}

We can now evaluate the spectral sum in Eq. (\ref{sum_kappa_spectrum}) to 
obtain an exact solution for the thermal conductivity. In particular,

\begin{widetext}
\be\label{Sum_kappa_exact}
S_{\text{$\kappa$}}(T) =
\frac{3}{\pi^2}\sum_{\ell=1}^{\infty}\,
\left[
\frac{1}{(\gamma+2\ell)(\gamma+2\ell-1)-2\lambda_{\kappa}}
\right]
\frac{\left|\sum_{p=1}^{\ell}\cC_{p}^{\ell}
      \cB(\frac{\gamma+1}{2},p+\onehalf)\right|^2}
     {\sum_{p=1}^{\ell}\sum_{q=1}^{\ell}
     \cC_{p}^{\ell}\cC_{q}^{\ell}\cB(\gamma+1,p+q-\onehalf)}
\,,
\ee
\end{widetext}
where 
\be\label{C_Gamma_ratio}
\cC_{p}^{\ell}\equiv C_{2p-1}^{2\ell-1}=(-4)^{p-1}
\frac{\Gamma(\gamma+\ell+p-\onehalf)}{\Gamma(2p)\Gamma(\ell-p+1)}
\,.
\ee

Although Eq. (\ref{Sum_kappa_exact}) provides us with an exact, closed form solution for the thermal
conductivity over the full temperature and pressure range, $T_c \le T \ll T_{f}$, the sums defining
$S_{\text{$\kappa$}}(T)$ involve ratios of Gamma functions. Thus, care must be taken in evaluating
these functions even for moderate values of their arguments. This is particularly true in the
low-temperature limit, $T\ll T_{\star}$, since the scaling parameter, $\gamma$ becomes large.
However, the limit $T\ll T_{\star}$ can also be evaluated using perturbation theory.


\subsection{Perturbation Theory}

At tempertures $T\ll T^{\star}$ inelastic quasiparticle collisions are relatively 
infrequent compared to elastic collisions off the aerogel. Thus, the inelastic 
collision integral in Eq. (\ref{linear_transport_equation}) is of order
\be
\frac{\delta I^{\text{in}}}{\delta I^{\text{el}}}
\sim 
\frac{\tau_{\text{el}}}{\tau_{\text{in}}}
\sim
\left(\frac{T}{T^{\star}}\right)^2 \equiv \delta
\ll 1
\,,
\ee
and we can formally expand the integral equation and the deviation from local 
equilibrium in the small parameter $\delta$, 
\be
\delta\bar{n}_{p_i}= \delta\bar{n}_{p_i}^{(0)}+ \delta\bar{n}_{p_i}^{(1)}+\ldots
\,.
\ee
The perturbation expansion through first order becomes,
\ber\label{zeroth_order}
L(\xi_{p_1},\hat\vp_1) 
&=& 
\delta I^{\text{el}}_{p_1}[\delta\bar{n}_{p}^{(0)}]
\,,
\\
\label{first_order}
0
&=& 
\delta I^{\text{in}}_{p_1}[\delta\bar{n}_{p}^{(0)}]
+
\delta I^{\text{el}}_{p_1}[\delta\bar{n}_{p}^{(1)}]
\,.
\eer
where $L(\xi_{p_1},\hat\vp_1)$ represents the driving term on the left side 
of Eq. (\ref{linear_transport_equation}). For heat transport the zeroth-order 
solution of Eq. (\ref{zeroth_order}) is simply the distribution in the elastic 
scattering limit, 
\be
\delta\bar{n}_{p_i}^{(0)}=
\left(\pder{n_0}{\xi_{p_i}}\right)
\left(\frac{\xi_{p_i}}{T}\right)
\left(\vv_{p_i}\cdot\grad T\right)
\tau_{\text{el}}
\,,
\ee
and now provides the driving term for the first order correction in 
Eq. (\ref{first_order}). This equation
has the same integral kernel as that of Eq. (\ref{zeroth_order}) and so 
we can express the first-order correction in terms of an inelastic 
correction to the scattering time,
\be
\delta\bar{n}_{p_i}^{(1)}=
\left(\pder{n_0}{\xi_{p_i}}\right)
\left(\frac{\xi_{p_i}}{T}\right)
\left(\vv_{p_i}\cdot\grad T\right)
\tau_{1}(\xi_{p_i})
\,,
\ee
where $\tau_1$ is the first-order correction to the mean scattering time, 
$\tau_{\text{el}}$. This distribution function gives the first order 
elastic collision integral,
\be
\delta I^{\text{el}}_{p_1}[\delta\bar{n}_{p}^{(1)}]
=
-\left(\pder{n_0}{\xi_{p_1}}\right)
\left(\frac{\xi_{p_1}}{T}\right)
\left(\vv_{p_1}\cdot\grad T\right)
\frac{\tau_{1}}{\tau_{\text{el}}}
\,.
\ee
The solution of the Eq. (\ref{first_order}) can then be expressed in terms of $\tau_1$. 
The analysis of the inelastic collision integral, {\sl evaluated} with the zeroth order 
nonequilibrium distribution function, $\delta\bar{n}_{p_i}^{(0)}$, leads to
\be\label{perturbative_tau1}
\begin{split}
\tau_1 =&-\frac{\tau_{\text{el}}^2}{\tau_{\text{in}}}
        (n_0(t_1)(1-n_0(t_1))\,t_1)^{-1}
\\
	&\times
	\Bigg[n_0(t_1)\,I_1\,t_1 +\lambda_{\kappa} n_0(t_1)\,J_1 \Bigg]
\,,
\end{split}
\ee
where
$I_1 = I(\xi_1)/T^2$, with $I(\xi_1)$ given by Eq. \ref{integral_I}, and
\ber
\label{integral_J1}
J_1 &=&\frac{1}{T^3}\int d\xi_2\,\xi_2\,n_0(\xi_2)\,K(\xi_1+\xi_2)
\nonumber\\
    &=& -\onesixth\left(\pi^2+t^2\right)\,t_1\left(1-n_0(t_1)\right)\,,
\,,
\eer
with $K(\xi)$ defined by Eq. \ref{integral_K}.
The resulting first-order correction for the collision time reduces to
\be
\tau_1 = -\onehalf\frac{\tau_{\text{el}}^2}{\tau_{\text{in}}}
          \left(\pi^2 + t_1^2\right) 
	  \left(1-\onethird\lambda_{\kappa}\right)
\,,
\ee
which vanishes in the ``ballistic limit'' for inelastic collisions, 
$\lambda_{\kappa}\rightarrow 3$.

The first order correction to the thermal conductivity is calculated by 
evaluating Eq. (\ref{quasiparticle_heat_current}) with the first-order 
correction, $\delta\bar{n}_{p}^{(1)}$. Writing
$\vj_{\text{q}}^{(1)}=-\delta\kappa\grad T$, we obtain,
\be
\delta\kappa = \twothirds N_f v_f^2\,T\,
\int_{-\infty}^{+\infty} dt\frac{t^2}{4\cosh^2(t/2)}\tau_1(t)
\,.
\ee

After the integration over $\tau_1(t)$, and scaling to the elastic limit for 
the thermal conductivity given
in Eq. (\ref{thermal_conductivity_elastic}), we obtain,
\be
\frac{\delta\kappa}{\kappa_{\text{el}}}=-\frac{6}{5}\pi^2
\left(1-\onethird\lambda_{\kappa}\right)
\left(\frac{\tau_{\text{el}}}{\tau_{\text{in}}}\right)
\,.
\ee
\section{Results}\label{Sec:Results}

\begin{figure}[h]
  \epsfysize=\hsize\rotate[r]{\epsfbox{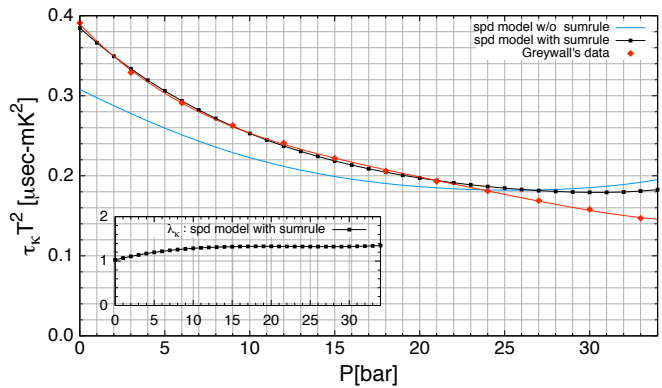}}
  \caption{Pressure dependence of the low-temperature limit of the thermal transport 
           time $\tau_{\kappa}T^2$ for pure \He. The \red{\bf red -$\blacklozenge$-} are the
        data from Ref. \onlinecite{gre84}. The \lightblue{\bf blue ---} is the \emph{spd} 
       model without enforcing the FSSR, and the [{\bf black $\blacksquare$}] are the \emph{spd}
       model with the FSSR enforced. Inset: $\lambda_{\kappa}$ calculated in the \emph{spd} 
       model with the FSSR enforced.}
  \label{fig-bulk_thermal_transport_time}
\end{figure}

Theoretical models for the quasiparticle collision probability, $W(\theta,\phi)$, for pure \He\ have
been proposed by a number of authors.\cite{dy69,sau81b,pfi83,lev83} We use an extended version of
the {\sl s-p model} introduced by Dy and Pethick\cite{dy69}, described as the {\sl spd} model in
Sec. \ref{appendix-scattering_model}. 
In Fig. \ref{fig-bulk_thermal_transport_time} we compare the results for the thermal transport
scattering time, $\tau_{\kappa} T^2$, for pure \He\ calculated in the {\sl spd} model with the
Landau parameters taken from Refs. \onlinecite{hal90,har00} and the limiting low-temperature thermal
conductivity measurements from Greywall (Table II of Ref. \onlinecite{gre84}).
The theoretical and experimental results are in agreement over the pressure range,
$p=0-25\,\mbox{bar}$ provided the forward-scattering sum rule (FSSR) is enforced (see Eq.
(\ref{FSSR}) in the Appendix). Above $25\,\mbox{bar}$ there are deviations
$0\%\le\delta\tau_{\kappa}/\tau_{\kappa}\le 14\%$ indicating the role of additional scattering not
described by the {\sl spd} model. Note in particular that the dimensionless scattering parameter,
$\lambda_{\kappa}$, is nearly constant over the entire pressure range, i.e. $1.0
\lesssim\lambda_{\kappa}\lesssim 1.3$.

\subsection{Results for \Heaero} 

\begin{figure}[h]
\hspace*{2mm}\epsfysize=3in{\epsfbox{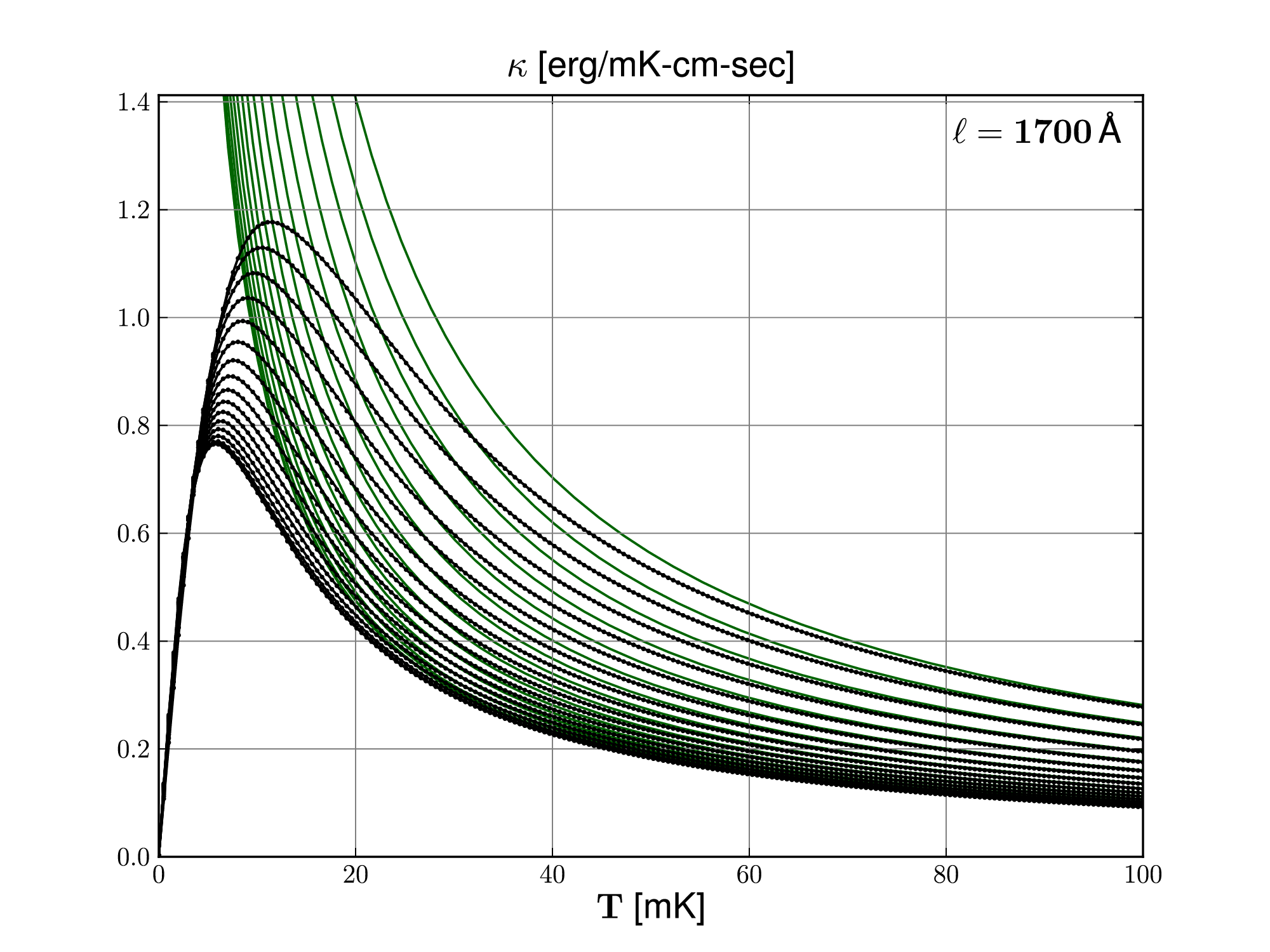}}
  \caption{Theoretical results for the thermal conductivity of \Heaero\ vs. 
		   $T$ and $p$ for an elastic \emph{mfp} of $\ell = 1700\,\mbox{\AA}$ are shown 
		   in \textbf{\textcolor{black}{black $-\bullet-$}}; results for pure \He\ are
		   shown as \darkgreen{green lines}. The pressure ranges from $p=0-32\,\mbox{bar}$ in
		   steps of $2\,\mbox{bar}$ starting with the the upper curve at $p=0\,\mbox{bar}$.
		  }
  \label{fig:Kappa_vs_T_and_p-3He-aero}
\end{figure}

Theoretical results for heat transport in \Heaero\ based on the two-channel solution for the thermal
conductivity are shown in Fig. \ref{fig:Kappa_vs_T_and_p-3He-aero}. In addition to the \emph{mfp}
describing the aerogel, the input data for bulk \He\ used to generate these results density ($n$),
effective mass ($m^{\star}$), Fermi velocity ($v_f$) and the Fermi liquid parameters ($F_{l}^{s,a}$)
for $l\le 2$, all of which are taken from the database provided in Ref. \onlinecite{har00} and
\onlinecite{hal90}. The Fermi-liquid parameters are used to construct the inelastic scattering rate
using the {\sl spd} model as described in the Appendix (Sec. \ref{appendix-scattering_model}).

\begin{figure}[h]
  \epsfysize=3in{\epsfbox{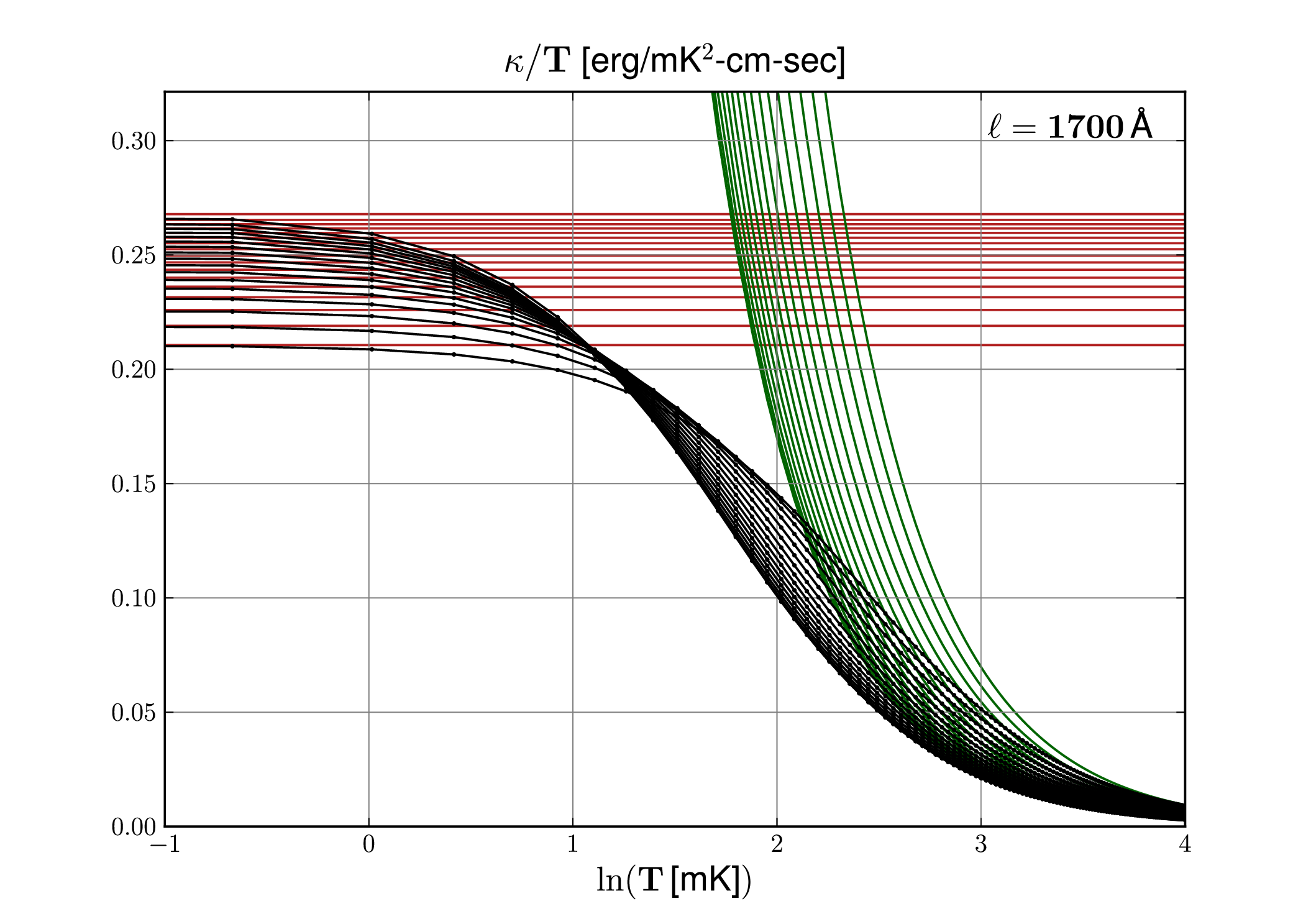}}
  \caption{$\kappa/T$ vs. $\ln(T[\mbox{mK}])$ and $p[\mbox{bar}]$. The inputs and labels are the
            as Fig. 
			\ref{fig:Kappa_vs_T_and_p-3He-aero}. The slopes for the elastic limit,
			 $\kappa_{\text{el}}/T$, are shown in \textbf{\textcolor{darkred}{darkred $-$}}.
           }
  \label{fig:Kappa/T_vs_lnT_and_p-3He-aero}
\end{figure}

The cross-over from the high-temperature regime dominated by inelastic quasiparticle collisions to
the low-temperature regime dominated by elastic scattering by the disordered medium occurs over a
fairly broad temperature range for dilute aerogels with long \emph{mfp}. The elastic regime below
$T\simeq 5-10\,\mbox{mK}$ is well described by $\kappa = \kappa_{\text{el}} + \cO(T^3)$, with
$\kappa_{\text{el}}$ given by Eq. (\ref{thermal_conductivity_elastic}). The pressure dependence of
the slope of $\kappa_{\text{el}}(T)$, while not visible in Fig. \ref{fig:Kappa_vs_T_and_p-3He-aero},
is shown clearly in Fig. \ref{fig:Kappa/T_vs_lnT_and_p-3He-aero}. Note that $\lim_{T\rightarrow
0}\kappa/T$ can provide a determination of the elastic \emph{mfp} for the aerogel.

\begin{figure}[h]
  \epsfysize=3in{\epsfbox{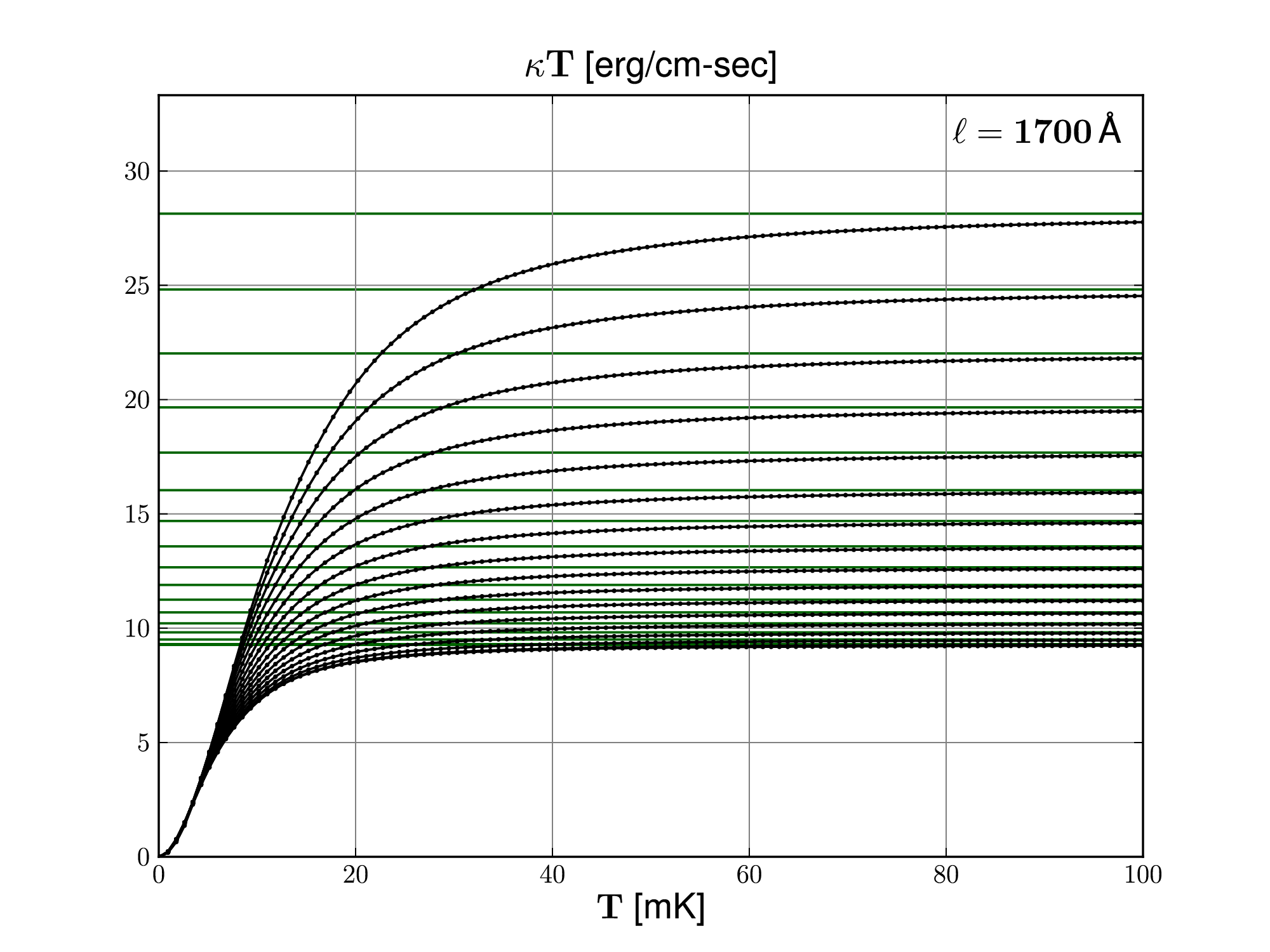}}
  \caption{$\kappa T$ vs. $T[\mbox{mK}]$ and $p[\mbox{bar}]$. The inputs and labels are the
            same as those of Fig. \ref{fig:Kappa_vs_T_and_p-3He-aero}. Note that results for
			$\kappa T$ for bulk \He\ are shown as \textbf{\textcolor{darkgreen}{darkgreen {---}}}.
           }
  \label{fig:KappaT_vs_T_and_p-3He-aero}
\end{figure}

Figure \ref{fig:Kappa/T_vs_lnT_and_p-3He-aero} for $\kappa/T$ highlights the
deviations in the thermal conductivity from the elastic limit limit even at temperatures of order a
few milli-Kelvin. Similarly, in the high temperature limit the product, $\kappa T$, approaches the
bulk \He\ limit determined by inelastic scattering. Significant deviations from the pure \He\ limit
are shown in Fig. \ref{fig:KappaT_vs_T_and_p-3He-aero} over a wide range of temperatures above
$T_{\star}$.

\subsection{Comparison with Experiments}

\begin{figure}[h]
\hspace*{2mm}\epsfysize=3.1in{\epsfbox{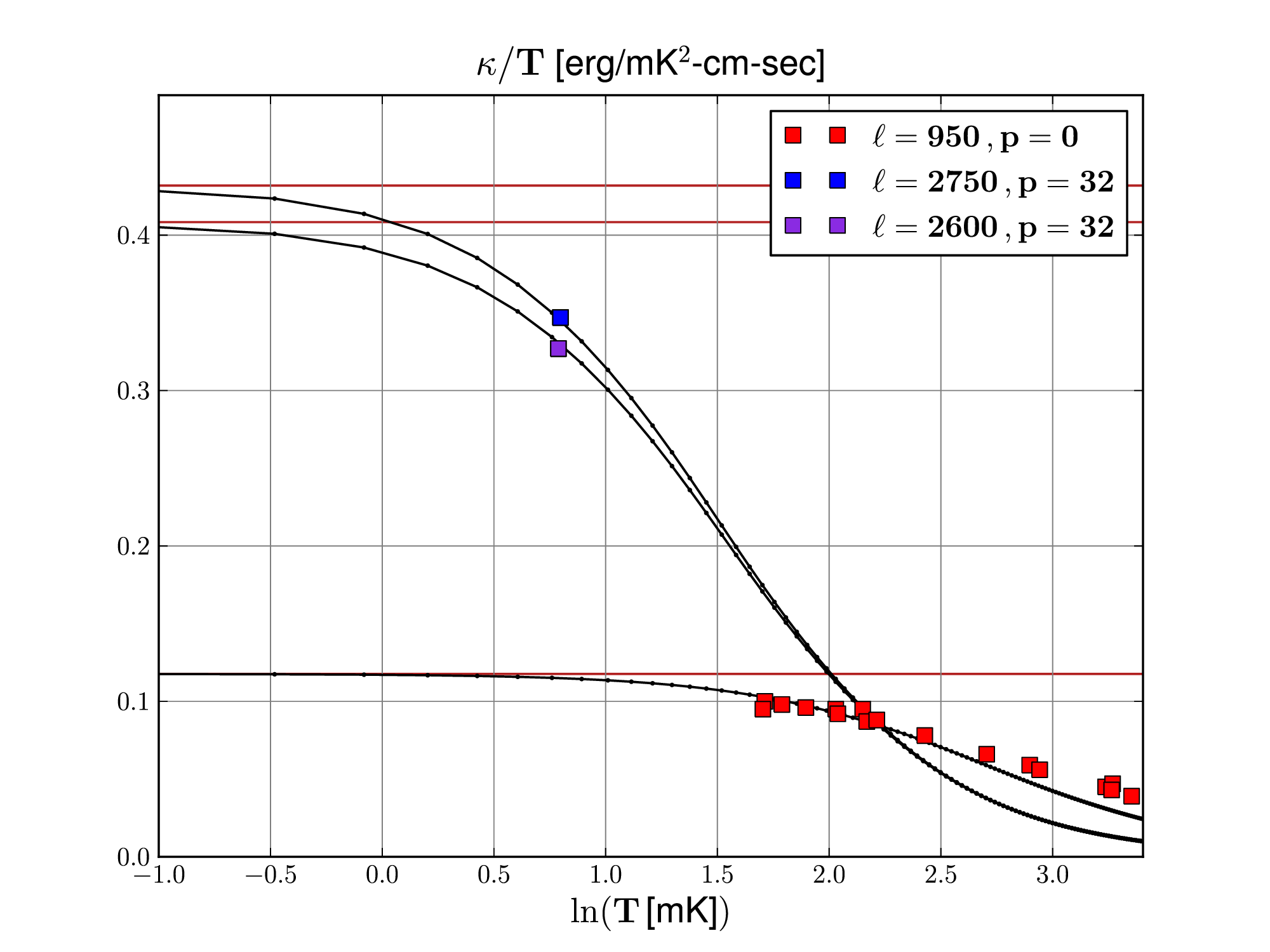}}
  \caption{Comparison $\kappa/T$ vs. $\ln(T[\mbox{mK}])$ with experiments. Data shown  
           as \textbf{\textcolor{red}{$\blacksquare$}} is from the Lancaster group 
           for a $98\%$ aerogel and a pressure of $p\simeq 0\,\mbox{bar}$.\cite{ree02} 
		   Data shown as \textbf{\textcolor{blue}{$\blacksquare$}} 
           (\textbf{\textcolor{purple}{purple $\blacksquare$}}) is from the Stanford group with two 
		   monolayers of \Hefour\ coating (without \Hefour) the silica strands at high pressure
		   ($p=32.4\,\mbox{bar}$),
           also for $98\%$ aerogel but grown in a different laboratory. The
 		   \textbf{\textcolor{darkred}{---}} show $\lim_{T\rightarrow 0}\kappa/T$.
           }
  \label{fig:Kappa/T_vs_lnT_and_p-3He-aero-Comparison}
\end{figure}

Barker {\it et.al}\cite{bar98} reported results for the thermal conductivity of \He\ in 98\% at
aerogel of $\kappa=7.2\,\mbox{ mW/mK}$ at $p = 32.4\,\mbox{bar}$ and $T= 2.20\,\mbox{mK}$. They also
added two monolayers for \Hefour\, which displaces the solid \He\ coating the silica aerogel
strands, and measured a slight \emph{increase} in the thermal conductivity, i.e. $\kappa=7.7\,\mbox{
mW/mK}$ at $T= 2.22\,\mbox{mK}$. Comparison of these two data points with the theoretical
predictions for this pressure are shown in Fig. \ref{fig:Kappa/T_vs_lnT_and_p-3He-aero-Comparison}.

If the difference in the \emph{mfp} with and without the \Hefour\ is attributable to spin-exchange
scattering of itinerant \He\ spins by the localized solid \He\ spins,\cite{bar00a,sau03} then we can
estimate the contribution to the scattering rate from indirect spin-exchange scattering to be,
\be
\frac{1}{\tau_{\text{spin}}} = v_f\left(\frac{1}{\ell_{\text{\He}}}
                                          -\frac{1}{\ell_{\text{\He+\Hefour}}} \right)
\,,
\ee
and thus a mean time for spin-exchange scattering of $\tau_{\text{spin}}\simeq
0.15\,\mu\mbox{sec}$, i.e. several orders of magnitude longer than the mean time for elastic
scattering off the aerogel strands, $\tau_{\text{el}}\simeq \ell_{\text{\He+\Hefour}}/v_f\simeq
8.6\,\mbox{ns}$. For scattering off a random distribution of $N_s$ localized spins via a Kondo
interaction,
\be
u=-\sum_{i=1}^{N_s}(J_{\text{ind}}/n)\vS_i\cdot\vsigma\,\delta(\vr-\vR_i)
\,,
\ee
the Born approximation implies an additional contribution to the scattering rate,
\be
\hbar/\tau_{\text{spin}} = \frac{4\pi n_s}{N_f}\left(J_{\text{ind}}/n\right)^2\,S(S+1)
\,.
\ee
Thus,we estimate the indirect exchange interaction to be 
\be
J_{\text{ind}} = 
E_f\left[\left(\frac{4}{9\pi}\right)
         \left(\frac{\hbar/\tau_{\text{spin}}}{E_f}\right)
         \left(\frac{n}{n_s}\right)\right]^{1/2}
\simeq 
0.5\,\mbox{mK}/\mbox{spin}
\,,
\ee
which is in agreement with the order of magnitude estimate for $J_{\text{ind}}$ inferred from the
absence of a low-field $A_1-A_2$ transition in \Heaero.\cite{sau03}

The Lancaster group also reports results for the low-temperature (i.e. $T\ll T_{\star}$) thermal
conductivity of normal \Heaero\ at low pressures, for aerogels with porosities of 95\% and
98\%.\cite{ree02} Results for $\varrho=98\%$ reported in Ref. \onlinecite{ree02} yield a
much smaller \emph{mfp}, $\ell = 950\,\AA$, than the Stanford data, suggesting significant
differences in aerogels of the same density prepared under different growth conditions. Note that
the authors of Ref. \onlinecite{ree02} attribute the deviations from the theoretical curve onsetting
near $T\simeq 20\,\mbox{mK}$ ($\ln(T[\mbox{mK}])\simeq 3\,$) to Kaptiza boundary conductance 
through the experimental cell walls. 

Although these results provide estimates for the aerogel \emph{mfp} they do not provide a test of
the the theory. Measurements of the thermal conductivity over the full temperature and pressure
range of normal \Heaero\ should provide a strong test of the two-channel theory based on homogeneous
disorder since we have an exact solution for the thermal conductivity in this model.
Conversely, if significant deviations from the theoretical predictions are observed they could
indicate new physics associated scattering and transport of fermionic excitations in a correlated
random medium. 

\begin{widetext}
\begin{center}
\begin{figure}[h]
\vspace*{.3cm}
\hspace*{0.5cm}\epsfysize=4in{\epsfbox{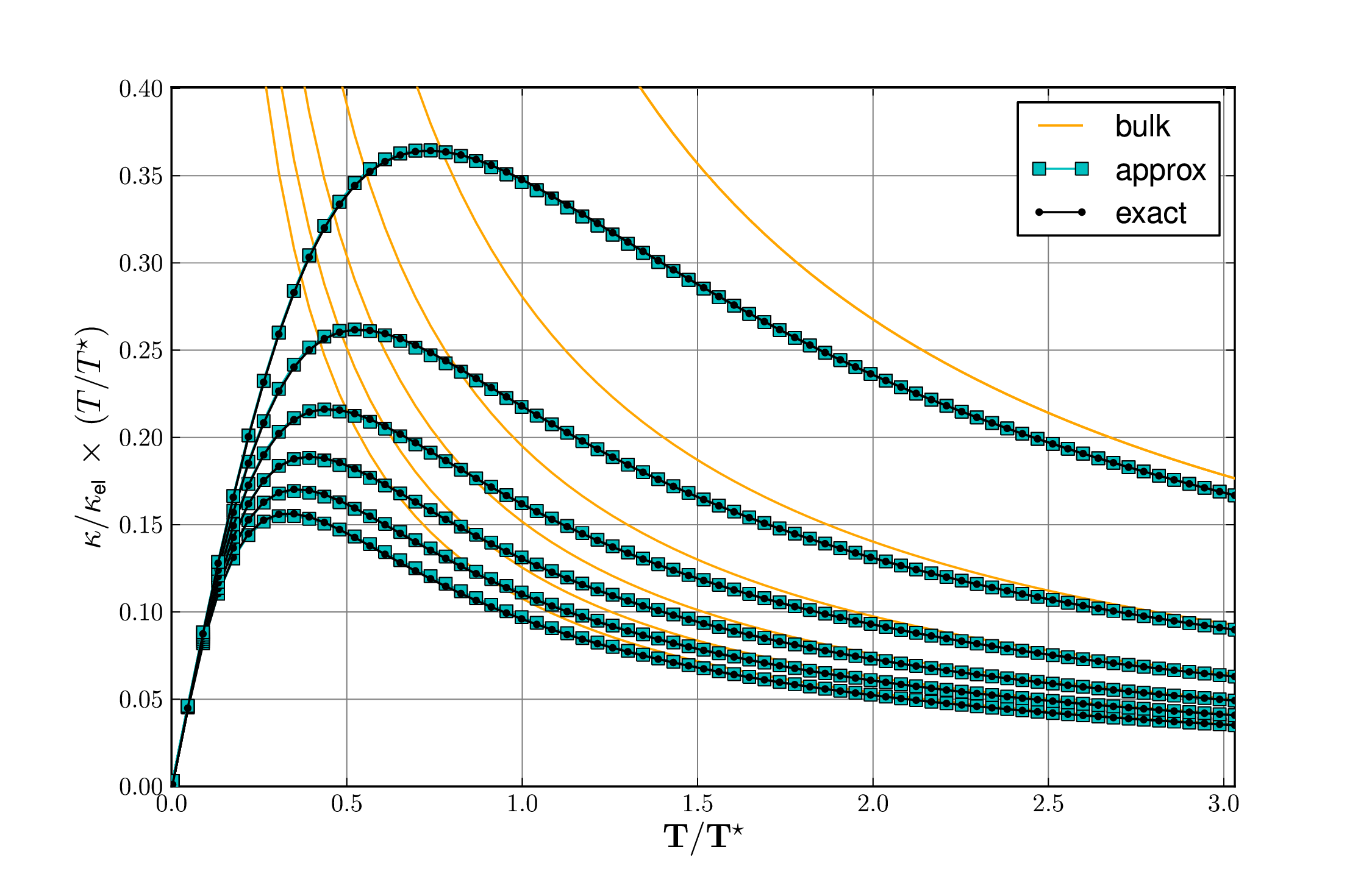}}
      \caption{Scaling function for the thermal conductivity plotted as 
           $x\,F(x,\lambda_{\kappa})\equiv\kappa/\kappa_{\text{el}}\times(T/T_{\star})$
           as a function of $x = T/T_{\star}$ for 
		   $\lambda_{\kappa}=0.0,\,0.5,\,1.0,\,1.5,\,2.0,\,2.5$, 
		   starting from the lowest to the highest curve, respectively. The exact results
		   based on Eqs. \ref{structure_constant_Tstar},\ref{C_Gamma_ratio} and 
		   \ref{Sum_kappa_exact} are shown as the \textbf{\textcolor{black}{Black $-\bullet-$}}, 
		   the \textbf{\textcolor{cyan}{Cyan $\blacksquare$}} are based on the approximate analytic 
		   formula given in Eqs. (\ref{approximate_scaling}, \ref{weight_function}, 
		   \ref{MR_scaling} and \ref{exponential_scaling}), and the 
		   \textbf{\textcolor{orange}{Orange $-$}} are 
		   the results for pure, bulk \He\ normalized to the elastic limit for \Heaero.           
		   }
     \label{fig:Kappa_scaling}
\end{figure}
\end{center}
\end{widetext}

\subsection{Scaling Function}
\vspace*{-0.5cm}
The exact solution for the thermal conductivity in the two-channel scattering theory for \Heaero\
can be expressed in terms of a scaling function. Normalizing $\kappa$ by the thermal conductivity in
the elastic limit, $\kappa_{\text{el}}$, from Eq. (\ref{thermal_conductivity_elastic}) gives,
\be\label{F_scaling}
\frac{\kappa}{\kappa_{\text{el}}} 
= 
\frac{\tau_{\text{in}}}{\tau_{\text{el}}}
\times S_{\text{$\kappa$}}(T)
\equiv
F(T/T^{\star},\lambda_{\kappa})
\,.
\ee
Note that $\tau_{\text{el}}/\tau_{\text{in}}\equiv(T/T^{\star})^2$, and that
$S_{\text{$\kappa$}}(T)$ calculated from Eq. (\ref{sum_kappa_spectrum}) [Eq. (\ref{Sum_kappa_exact})
in Sec. \ref{sec:exact_solution}] provides the \emph{exact} scaling function, $F(x,\lambda)$, since
$S_{\text{$\kappa$}}(T)$ depends only on $x=T/T^{\star}$ and the scattering ratio, 
$-1 <\lambda_{\kappa}< 3$.
\emph{Thus, the test of the two-channel transport theory would be to demonstrate that the thermal
conductivity of \Heaero\ obeys the scaling behavior over the full temperature and pressure range of
the normal state, \underline{and} a broad range of aerogel density and mfp.
}

The exact solution for $\kappa/\kappa_{\text{el}}\times (T/T_{\star})\equiv x\,
F(x,\lambda_{\kappa})$ is shown in Fig. (\ref{fig:Kappa_scaling}). The calculation of spectral sum,
$S_{\text{$\kappa$}}(T)$, was carried out using arbitrary-precision floating point arithmetic in
order to evaluate the ratios of the Gamma functions or large arguments that enter Eq.
(\ref{Sum_kappa_exact}) with sufficient precision to obtain accurate results for the triple sum that
defines $S_{\text{$\kappa$}}(T)$. In particular, the points labeled ``exact'' in Fig.
(\ref{fig:Kappa_scaling}) were obtained with the floating point precision set at $55$ digits and
each sum was cutoff after $30$ terms were computed. One can obtain reasonably good results with a
lower precision setting for the floating point arithmetic, but double precision on a 32-bit
machine limits the accuracy of the results, particularly in the limit $T <
T_{\star}$.

Also shown in Fig. (\ref{fig:Kappa_scaling}) are calculations of the scaling function based on an
approximate analytic formula that is numerically fast and easy to evaluate. The approximate scaling
function is constructed from the the asymptotic limits for $S_{\text{$\kappa$}}(T)$ for $T\gg
T_{\star}$ and $T\rightarrow 0$, as well as the leading order perturbative result for $T\ll
T_{\star}$, as described below.

The limiting behavior for the exact scaling function is known from the asymptotic limit, $x\gg 1$,
and perturbation theory about $x=0$. In particular,
\be\label{scaling_exact_limits}
F(x,\lambda_{\kappa}) =
\Bigg\{
\begin{array}{ll}
1 - \frac{6}{5}\pi^2(1-\tinyonethird\lambda_{\kappa})\,x^2 &\,, x \ll 1
\cr
\displaystyle{\frac{1}{x^2}\,S_{\text{$\kappa$}}^{\text{$\infty$}}} &\,, x \gg 1\,,
\end{array}
\ee
where $S_{\text{$\kappa$}}^{\text{$\infty$}}$ is given by Eq. (\ref{Sum_kappa_pure}).

\vspace*{1cm}
\begin{figure}[!]
  \epsfysize=0.7\hsize{\epsfbox{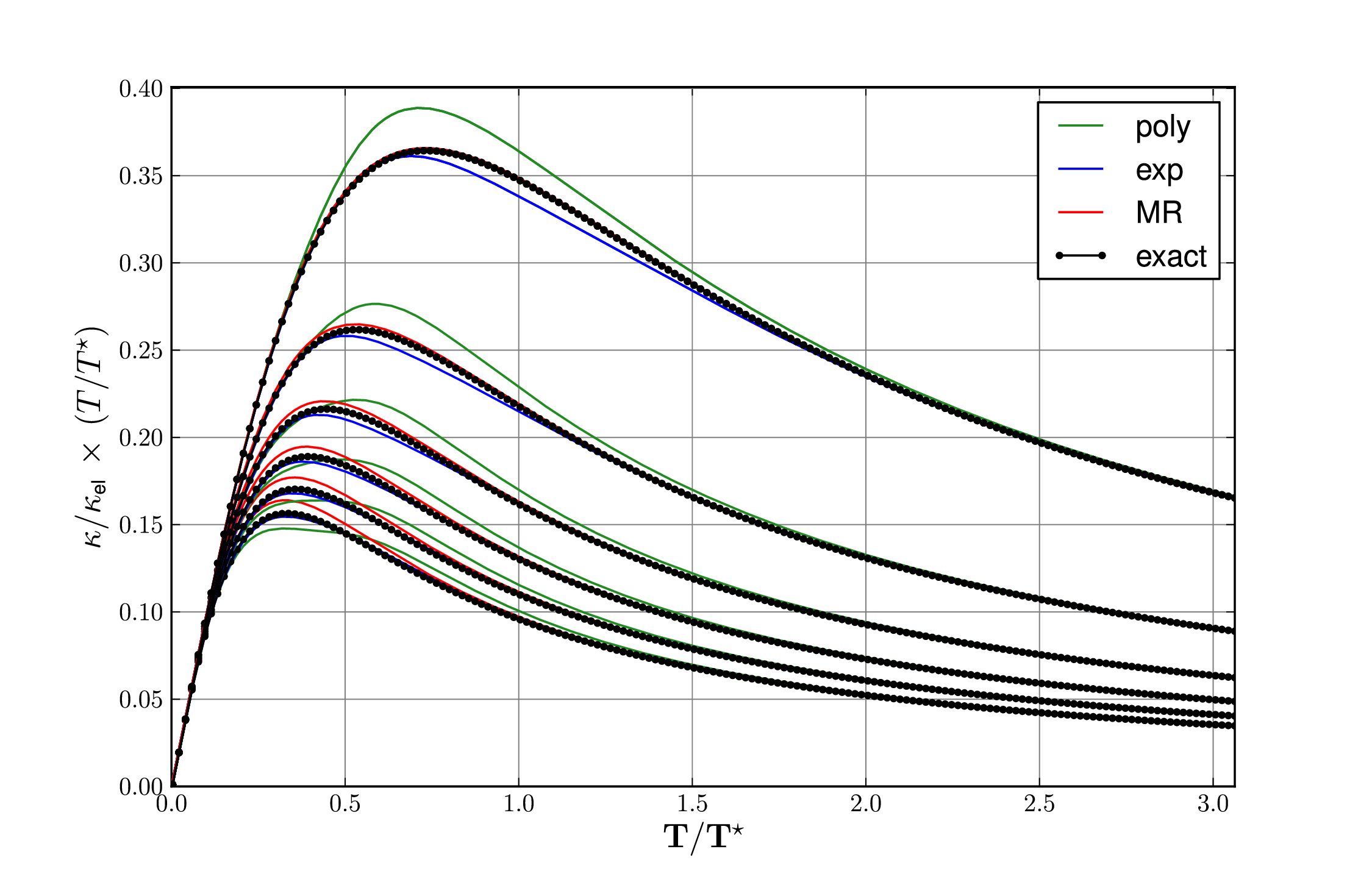}}
  \caption{Comparison of several approximates with the exact scaling function,
           $x\,F(x,\lambda_{\kappa})$, plotted as a function of $x = T/T_{\star}$ for 
		   $\lambda_{\kappa}=0.0,\,0.5,\,1.0,\,1.5,\,2.0,\,2.5$, 
		   starting from the lowest to the highest curve, respectively. The exact results
		   are shown in \textbf{\textcolor{black}{Black $-\bullet-$}}, while the approximate 
		   scaling functions, labelled as 
		   \textbf{$\mbox{\darkgreen{poly}}$},
		   \textbf{$\mbox{\blue{exp}}$} and
		   \textbf{$\mbox{\red{MR}}$}, are described in the text.           
		   }
  \label{fig:Kappa_scaling_compare}
\end{figure}

The most common approximate solution for multi-channel scattering is based \emph{Matthiessen's
Rule}, which in this context can be expressed as
\be
\frac{1}{\tau_{\kappa}^{\text{MR}}}
=
\frac{1}{\tau_{\text{el}}}
+
\frac{1}{S_{\text{$\kappa$}}^{\text{$\infty$}}\tau_{\text{in}}(T)}
\,,
\ee
i.e. the total transport scattering rate is the sum of independent rates for
purely elastic and purely bulk inelastic transport. The resulting expression for the the thermal
conductivity, normalized to its value in the elastic limit,
\be
\bar{\kappa}^{\text{MR}} 
=
\frac{\tau_{\text{$\kappa$}}^{\text{MR}}}{\tau_{\text{el}}}
\equiv
F_{\text{MR}}(T/T^{\star},\lambda_{\kappa})
\,,
\ee
defines the approximate scaling function, $F_{\text{MR}}(x,\lambda_{\kappa})$ given by,
\be\label{MR_scaling}
F_{\text{MR}}(x,\lambda_{\kappa}) = 
\frac{1}{1 + (S_{\text{$\kappa$}}^{\text{$\infty$}})^{-1}\,x^2}
=
\Bigg\{
\begin{array}{ll}
1-(S_{\text{$\kappa$}}^{\text{$\infty$}})^{-1}\,x^2
&\,, x\ll 1
\\
\displaystyle{S_{\text{$\kappa$}}^{\text{$\infty$}}\,\frac{1}{x^2}}
&\,, x\gg 1\,.
\end{array}
\ee

The MR scaling function deviates from the exact result of Eq. (\ref{scaling_exact_limits}) for the
leading order finite temperature correction. Curiously, the exact result for the leading order
correction is equal to that obtained from $F_{\text{MR}}(x,\lambda_{\kappa})$ by approximating
$S_{\text{$\kappa$}}^{\text{$\infty$}}$ with just the first term of the sum in Eq.
(\ref{Sum_kappa_pure}). This approximation is very good in the limit of nearly forward scattering.
In this limit the inelastic channel leads to large thermal transport for $T\gtrsim T_{\star}$. As a
result the MR scaling function gives a very good approximation to the exact scaling function in the
limit of large $\lambda_{\kappa}$ for all $x$. This is shown clearly in Fig.
(\ref{fig:Kappa_scaling_compare}). However, the MR scaling function deviates from the exact scaling
function when backscattering in the inelastic channel is significant, i.e. for $\lambda_{\kappa}
\lesssim 1.0$. These deviations are also clearly visible in Fig. (\ref{fig:Kappa_scaling_compare}).

We can try to improve on the MR scaling function by incorporating the exact perturbative result for
$F(x,\lambda_{\kappa})$ for $x \ll 1$. We construct an interpolation formula that connects the exact
asymptotic limits. A simple extension of Matthiessen's interpolation formula is the two-parameter,
rational polynomial function,
\be\label{polynomial_scaling}
F_{\text{poly}} = \frac{1}{1+a x^2} + \frac{x^2}{1+b x^4}
\,,
\ee
which has the limiting forms,
\be
F_{\text{poly}} =
\Bigg\{
\begin{array}{ll} 
1-(a-1)x^2 & \,, x \ll 1
\cr
\displaystyle{\left(\frac{1}{a}+\frac{1}{b}\right)\frac{1}{x^2}} & \,, x\gg 1
\,.
\end{array}
\ee
We then fix the coefficients from the exact asymptotic limits for $F(x,\lambda_{\kappa})$ in 
Eq. (\ref{scaling_exact_limits}),
\ber
a &=& 1 + \frac{6}{5}\pi^2\left(1 - \onethird\lambda_{\kappa}\right)
\\
b &=& \frac{a}{a\,S_{\kappa}^{\text{$\infty$}}-1}
\,.
\eer
Although this approximate scaling function works well for the $x\ll 1$, it does a poor job in the
intermediate and high-temperature region $x\gtrsim 1$ (green curves in Fig.
\ref{fig:Kappa_scaling_compare}), and is particularly poor for $\lambda_{\kappa} \rightarrow 3$.
If we consider the leading order correction to the asymptotic limit $x\rightarrow\infty$ we obtain
\be
F\rightarrow \frac{S_{\kappa}^{\infty}}{x^2} + C_{4}\,\frac{1}{x^4} + \cO(\frac{1}{x^6})
\,.
\ee
For the polynomial approximate we obtain,
\be
C_{4}^{\text{poly}} = -\frac{1}{(1+\frac{6\pi^2}{5}(1-\lambda_{\kappa}/3))^2}
\,,
\ee
while the  MR scaling function gives
\be
C_{4}^{\text{MR}} = -(S_{\kappa}^{\infty})^2
\,.
\ee
Both approximate scaling functions give the correct sign for the leading order correction, however
in the limit $\lambda_{\kappa}\rightarrow 3$, where we know the MR scaling function approaches the
exact result, we see that $C_{4}^{\text{MR}}$ is large and negative,
\be
C_{4}^{\text{MR}} \rightarrow -\left(\frac{5}{6\pi^2}\right)^2\frac{1}{(1-\lambda_{\kappa}/3)^2}
\,,
\ee
whereas $C_{4}^{\text{poly}} \rightarrow -1$. This discrepancy in $F_{\text{poly}}$ is traced to
the contamination of the temperature region $x >1$ by the exact solution that is valid for
$x\ll 1$.

We might remedy this problem with a two-parameter interpolation that limits the contamination
between $x\ll 1$ and $x\gg 1$. In particular, consider the approximate scaling function,
\be\label{exponential_scaling}
F_{\text{exp}} = \onehalf e^{-2ax^2} + \onehalf \left(1 - e^{-2b/x^2}\right)
\,.
\ee
For $x\ll 1$,
\ber
F_{\text{exp}}
             &\xrightarrow[x\ll 1]{}& 1 - ax^2 +  \cO(x^4)
\,.
\eer
Note that there are only exponentially small corrections to the leading order result for $x\ll
1$ coming from the terms that are fixed by the asymptotic solution for $x\gg 1$. Similarly, for
$x\gg 1$, the term that is fixed by the exact solution for $x\ll 1$ is now exponentially small and
we obtain,
\ber
F_{\text{exp}}  &\xrightarrow[x\gg 1]{}& 
\frac{b}{x^2}-\frac{b^2}{x^4}+\cO\left(\frac{1}{x^6}\right)
\,.
\eer
Using these expansions and the exact leading order asymptotic limits we obtain
\ber
a &=& \frac{6\pi^2}{5}(1-\lambda_{\kappa}/3)
\\
b &=& S_{\kappa}^{\infty}
\,.
\eer
This two-parameter interpolation formula yields a better approximation to the exact scaling
function, particularly for $\lambda_{\kappa} \lesssim 1$. However, $F_{\text{exp}}$ under estimates
the maximum in $x\,F(x,\lambda_{\kappa})$, and this deviation is enhanced as
$\lambda_{\kappa}\rightarrow 3$, as is clear from Fig. (\ref{fig:Kappa_scaling_compare}).
The basic result of this analysis is that the MR scaling function, $F_{\text{MR}}$, is accurate in
the limit of large $\lambda_{\kappa}$, but deviates from exact scaling for $\lambda_{\kappa}
\lesssim 1$. By contrast the two-parameter exponential scaling function, $F_{\text{exp}}$, is
accurate in limit $\lambda_{\kappa} < 1$, but shows increasing errors from exact scaling in the
cross-over region, $x\sim\cO(1)$, for $1 < \lambda_{\kappa} < 3$. This suggests that we combine
these two scaling functions into a single scaling function by weighting the respective regions of
accurate scaling, i.e.
\ber\label{approximate_scaling}
F_{\text{approx}}(x,\lambda_{\kappa}) 
&=& 
\mathsf{p}(\lambda_{\kappa})\,F_{\text{exp}}(x,\lambda_{\kappa})
\nonumber\\
&+&
\left(1-\mathsf{p}(\lambda_{\kappa})\right)\,F_{\text{MR}}(x,\lambda_{\kappa})
\,,
\eer

\noindent where the weight function $\mathsf{p}(\lambda_{\kappa})$ is chosen on the physical domain, 
$-1 < \lambda_{\kappa} < 3$, to satisfy, $\mathsf{p}(-1) = 1$, $\mathsf{p}(+3) = 0$.
Thus, the simplest weight functions which map the physical domain onto the interval $[0,1]$ are
\be\label{weight_function}
\mathsf{p}(\lambda_{\kappa}) = \left(\frac{1+\lambda_{\kappa}}{4}\right)^s
\,.
\ee
The quadratic weight function, i.e. $s=2$, leads to remarkably good agreement with the exact scaling
function for the entire domain of $\lambda_{\kappa}$ and reduced temperature,
$x=T/T_{\star}$. This comparison is shown in Fig. (\ref{fig:Kappa_scaling}). Note that the maximum
deviation for any of the computed values is less than $0.6\%$, and careful examination shows
that these small errors occur near the maxima of $x\,F(x,\lambda_{\kappa})$. Thus, the main
result here is that Eqs. \ref{MR_scaling}, \ref{exponential_scaling}, \ref{approximate_scaling} and
\ref{weight_function} provide numerically fast and accurate formulas for calculating the thermal
conductivity over the full temperature and pressure range within the two-channel scattering theory
for normal \Heaero.

\subsection{Scaling for \Heaero}

The analysis of the pressure dependence of the thermal conductivity of pure \He\ based on the spd
scattering amplitude described in Sec. \ref{Sec:Results} and App. \ref{appendix-scattering_model}
implies that the thermal transport scattering parameter is nearly pressure independent, i.e.
$\lambda_{\kappa}\simeq 1.3$ for $5\,\mbox{bar} \lesssim p \le 34\,\mbox{bar}$ with a smooth drop to
$\lambda_{\kappa} \simeq 1.0$ as pressures between $5$ and $0\,\mbox{bar}$ (see inset of Fig.
\ref{fig-bulk_thermal_transport_time}). 

\begin{figure}[h]
  \epsfysize=2.5in{\epsfbox{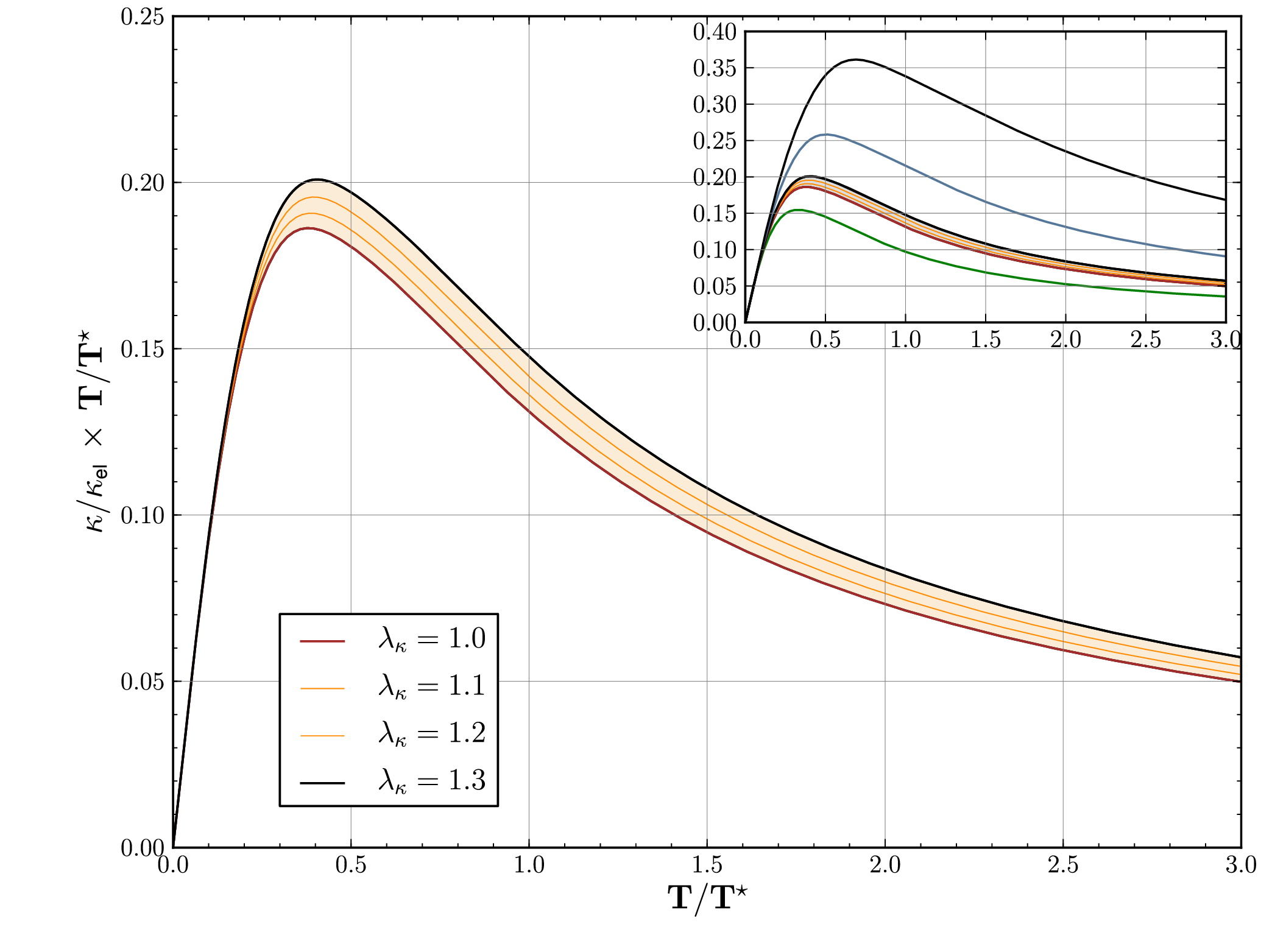}}
  \caption{Scaling of the thermal conductivity with $T/T_{\star}$ for \Heaero\ for
 			the physical range: $1.0 \le \lambda_{\kappa} \le 1.3$. Inset: Scaling 
			function for a much larger parameter range of transport scattering parameters, including
			$\lambda_{\kappa}=0.0, 2.0, 2.5$, in addition to the physical range.
           }
  \label{fig:Kappa_scaling_3He-aero}
\end{figure}

Pressure independence of the scattering parameter, $\lambda_{\kappa}$, implies that the thermal
conductivity for \emph{all temperatures} above the superfluid transition, \emph{all
pressures} and \emph{all elastic mean-free paths} should collapse to a single scaling function when
normalized to its value in the elastic scattering limit, i.e. $\lim_{T\rightarrow 0}\kappa =
\kappa_{\text{el}}$ given in Eq. (\ref{thermal_conductivity_elastic}). 
Thus, for \Heaero\ we expect that thermal conductivity for all $T$, $p$ and $\ell$ to collapse to
the narrow band of scaling functions shown in Fig. \ref{fig:Kappa_scaling_3He-aero}. A complete set
of measurements of the thermal conductivity of \Heaero\ for all $T$, $p$ and a wide range of aerogel
\emph{mfp} would provide a strong test of this theory, particularly the assumption of uncorrelated
disorder described by a single $mfp$. 

\medskip

\section{Summary}

Liquid \He\ impregnated into silica aerogel is a model system for investigating the effects of
quenched disorder on the properties of a strongly correlated Fermi liquid.
In the normal Fermi liquid the transport of heat, mass and spin by fermionic excitations exhibits
cross-over behavior from a high temperature regime, where inelastic scattering dominates, to a low
temperature regime dominated by elastic scattering off the aerogel.
The exact solution to the two-channel Boltzmann-Landau transport equation reported here provides
quantitative predictions for heat transport in liquid \Heaero. An approximate solution derived from
the asymptotic solutions and perturbation theory is accurate to less than $0.6\,\%$. 
A key result of this work is the scaling function, $F(T/T^{\star},\lambda_{\kappa})$,
that describes the exact solution for the normalized thermal conductivity,
$\kappa/\kappa_{\text{el}}$, for all pressures, temperatures (above $T_c$) and aerogel density.
A complete set of measurements of the thermal conductivity of \Heaero\ for all $T$, $p$ and a wide
range of aerogel \emph{mfp} would provide a strong test of this theory, particularly the predicted
scaling behavior based on two-channel scattering and the assumption of homogeneous disorder
described by a single $mfp$.
Conversely, systematic deviations from the predicted scaling function behavior should provide a
quantitative measure of the role of fractal correlations associated with the structure of the
aerogel.
The limited data that is available already hints that two-channel scattering is insufficient and
that spin-exchange scattering between itinerant \He\ spins and localized \He\ spins contributes to
the low-temperature thermal conductivity.
\subsubsection*{Acknowledgements}
This work was supported in part by National Science Foundation Grant DMR-0805277 (JAS) and the
Leverhulme Trust of the United Kingdom (PS).

\section{Appendix: \He\ Scattering Amplitude}\label{appendix-scattering_model} 

The binary collision amplitude for quasiparticles in pure \He\ in 
the low-energy region near the Fermi surface depends on
the momenta and the spin state of the initial and final pair of excitations. 
In particular the dimensionless scattering amplitude is

\begin{fmffile}{fmf_Tmatrix}
\be
\Tm_{\alpha_1\alpha_2;\alpha_3\alpha_4}(\vp_1,\vp_2;\vp_3,\vp_4)
=2N_f\times
\parbox{60mm}{
\begin{fmfgraph*}(60,40)
\fmftop{o1,o2}
\fmfstraight
\fmfbottom{i,f}
\fmfpolyn{empty,label=$\tm$}{v}{4}
\fmf{fermion}{i,v1}
\fmf{fermion}{f,v2}
\fmf{fermion}{v4,o1}
\fmf{fermion}{v3,o2}
\fmflabel{$\vp_1\alpha_1$}{i}
\fmflabel{$\vp_2\alpha_2$}{f}
\fmflabel{$\vp_3\alpha_3$}{o1}
\fmflabel{$\vp_4\alpha_4$}{o2}
\end{fmfgraph*}
}
\ee
\end{fmffile}

\noindent 
where $\tm$ is formally defined by the matrix elements of a transition operator 
between incoming $(1,2)$ and outgoing $(3,4)$ quasiparticles.
For a Fermi liquid with only exchange interactions such as \He\ the total spin 
$S$ and any one component, $S_z$, are conserved by collisions. For an unpolarized 
Fermi liquid there is no preferred direction for the spins to align. As a result 
all three spin-triplet amplitudes are equal and there are only two independent 
amplitudes corresponding to the total spin $S=0$ and $S=1$, which we label as 
the singlet ($s$) and triplet ($t$) amplitudes,
\ber
\hspace*{-8mm}
\Tm_{s} &=& \onehalf\left[
                \Tm_{\uparrow\downarrow;\uparrow\downarrow} -
                \Tm_{\uparrow\downarrow;\downarrow\uparrow} -
                \Tm_{\downarrow\uparrow;\uparrow\downarrow} +
                \Tm_{\downarrow\uparrow;\downarrow\uparrow}
                        \right]                                 \\
\Tm_{t} &=&  \onehalf\left[
                \Tm_{\uparrow\downarrow;\uparrow\downarrow} +
                \Tm_{\uparrow\downarrow;\downarrow\uparrow} +
                \Tm_{\downarrow\uparrow;\uparrow\downarrow} +
                \Tm_{\downarrow\uparrow;\downarrow\uparrow}
                        \right]
\nonumber\\
             &=& \Tm_{\uparrow\uparrow;\uparrow\uparrow}
             = \Tm_{\downarrow\downarrow;\downarrow\downarrow}
\,.
\eer
Also note that amplitudes which differ by inversion of all the 
spin projections are equal,
\be
\Tm_{\uparrow\uparrow;\uparrow\uparrow} = 
\Tm_{\downarrow\downarrow;\downarrow\downarrow}
\,,\quad
\Tm_{\uparrow\downarrow;\uparrow\downarrow} = 
\Tm_{\downarrow\uparrow;\downarrow\uparrow}
\,,\quad
\Tm_{\uparrow\downarrow;\downarrow\uparrow} = 
\Tm_{\downarrow\uparrow;\uparrow\downarrow}
\,.
\ee
Thus, we use a short-hand notation,
\ber
\Tm_{\uparrow\uparrow}&\equiv& \Tm_{\uparrow\uparrow;\uparrow\uparrow}
              = \Tm_{\downarrow\downarrow;\downarrow\downarrow} \\
\Tm_{\uparrow\downarrow}&\equiv& \Tm_{\uparrow\downarrow;\uparrow\downarrow}
              = \Tm_{\downarrow\uparrow;\downarrow\uparrow}     \\
\tilde\Tm_{\uparrow\downarrow} &\equiv& \Tm_{\uparrow\downarrow;\downarrow\uparrow}
              = \Tm_{\downarrow\uparrow;\uparrow\downarrow}
\,,
\eer
and express the spin-projection amplitudes in terms of the singlet and 
triplet amplitudes
\ber
\Tm_{\uparrow\uparrow}         &=& \Tm_{t}                                  \\
\Tm_{\uparrow\downarrow}       &=& \onehalf\left(\Tm_{t}+\Tm_{s}\right) \\
\tilde\Tm_{\uparrow\downarrow} &=& \onehalf\left(\Tm_{t}-\Tm_{s}\right)
\,.
\eer

The $\Tm$-matrix can then be expressed in terms of $\Tm_{t,s}$ and the 
corresponding symmetric (triplet) and anti-symmetric (singlet) 
spin matrix elements,
\be\label{t-matrix_singlet-triplet}
\Tm_{\alpha_1\alpha_2;\alpha_3\alpha_4} = 
\Tm_{t}\,
\Sig^{(+)}_{\alpha_1\alpha_2;\alpha_3\alpha_4}
+\Tm_{s}\,
\Sig^{(-)}_{\alpha_1\alpha_2;\alpha_3\alpha_4}
\,,
\ee
where 
\be
\Sig^{(\pm)}_{\alpha_1\alpha_2;\alpha_3\alpha_4}
=
\onehalf\left(
\delta_{\alpha_1\alpha_3}\delta_{\alpha_2\alpha_4}
\pm
\delta_{\alpha_1\alpha_4}\delta_{\alpha_2\alpha_3}
\right)
\,.
\ee
Since there are only two independent amplitudes it is often useful to use 
the \emph{symmetric} and \emph{anti-symmetric} amplitudes defined as
\ber
\Tm^{s}&=&\onehalf\left(\Tm_{\uparrow\uparrow}+\Tm_{\uparrow\downarrow}\right)
        = \onefourth\left(3\Tm_{t}+\Tm_{s}\right) \\
\Tm^{a}&=&\onehalf\left(\Tm_{\uparrow\uparrow}-\Tm_{\uparrow\downarrow}\right)
        = \onefourth\left(\Tm_{t}-\Tm_{s}\right)
\,.
\eer

Inverting, we have
\ber\label{triplet-ph_amplitudes}
\Tm_{t}&=&\left(\Tm^{s}+\Tm^{a}\right) 
\\
\Tm_{s}&=&\left(\Tm^{s}-3\Tm^{a}\right) 
\,.
\eer
The two sets of amplitudes, $\Tm^{s,a}$ or $\Tm_{t,s}$, define different, 
but equivalent representations for the spin-dependent $\Tm$ matrix. 
The $\Tm^{s,a}$ amplitudes are the amplitudes for the $\Tm$-matrix expressed 
in terms of the direct "particle-hole" channel, $1\rightarrow 3$ and $2\rightarrow 4$,
\be
\Tm_{\alpha_1\alpha_2;\alpha_3\alpha_4}
=
\Tm^{s}\,\delta_{\alpha_1\alpha_3}\,\delta_{\alpha_2\alpha_4}
+
\Tm^{a}\,\vsigma_{\alpha_1\alpha_3}\cdot\vsigma_{\alpha_2\alpha_4}
\,.
\ee

For quasiparticle scattering on the Fermi surface the
scattering amplitudes, $\Tm_{t,s}$, reduce to functions of the {\sl directions} 
of the quasiparticle momenta on the Fermi surface,
\be
\Tm_{t,s}(\vp_1,\vp_2;\vp_3,\vp_4) 
\leadsto 
\Tm_{t,s}(\hat\vp_1,\hat\vp_2;\hat\vp_3,\hat\vp_4)
\,.
\ee
Furthermore, rotational invariance implies that $\Tm_{t,s}$ can be expressed in 
terms of the {\sl relative} direction cosines,
\ber\label{direction_cosines}
x_2 &=& \hat\vp_2\cdot\hat\vp_1 \equiv \cos\theta\,\,
     = \hat\vp_3\cdot\hat\vp_4\,,
\\
x_3 &=& \hat\vp_3\cdot\hat\vp_1 = \cos\theta_3 
     = \hat\vp_4\cdot\hat\vp_2\,,
\\
x_4 &=& \hat\vp_4\cdot\hat\vp_1 = \cos\theta_4 
     = \hat\vp_3\cdot\hat\vp_3\,.
\eer
The fourth column of equalities follows from momentum conservation 
for $|\vp_i|=p_f$,
\be
\hat\vp_1+\hat\vp_2 = \hat\vp_3+\hat\vp_4
\,.
\ee
The conservation law also implies that there are only {\sl two} independent 
angles. We adopt Abrikosov and Khalatnikov's parametrization \cite{abr58} 
in terms of the angle $\theta$ between the two incoming momenta, and $\phi$, 
the angle between the planes defined by $\vn=\hat\vp_1\times\hat\vp_2$ and 
$\vn'=\hat\vp_3\times\hat\vp_4$,
\be
\cos\phi = \frac{\vn\cdot\vn'}{|\vn|\,|\vn'|} 
=\frac{x_3-x_4}{1-x_2}
\,.
\ee
Thus, $\Tm_{t,s}(\hat\vp_1,\hat\vp_2;\hat\vp_3,\hat\vp_4) 
=\Tm_{t,s}(\theta,\phi)$.

The Pauli exclusion principle requires the $\Tm$-matrix to be 
anti-symmetric under exchange of either the initial or the final state 
of the two fermions.
Thus, the spin-singlet (triplet) amplitude is necessarily symmetric 
(anti-symmetric) under exchange of the initial or final momenta, or in 
terms of the scattering angle,
\ber
\Tm_{s}(\theta,\phi+\pi) &=& +\Tm_{s}(\theta,\phi)
\\
\Tm_{t}(\theta,\phi+\pi) &=& -\Tm_{t}(\theta,\phi)
\,.
\eer
Thus, we can formally expand the singlet (triplet) amplitudes as a sum 
over even (odd) functions of $\cos(m\phi)$,
\ber\label{T_singlet_expansion}
\Tm_{s}(\theta,\phi) 
&=& 
\sum_{m=0}^{\text{even}}\,
A_{s}^{(m)}(\cos\theta)\,\cos(m\phi)
\,,
\\
\Tm_{t}(\theta,\phi) 
&=& 
\sum_{m=1}^{\text{odd}}\,
A_{t}^{(m)}(\cos\theta)\,\cos(m\phi)
\,.
\label{T_triplet_expansion}
\eer
Note that $\Tm_{t}$ vanishes for $\phi=\pi/2$ and $\phi=3\pi/2$. For these 
angles the momentum transfer in the direct and exchange channels is identical, 
in which case exchange symmetry requires the triplet amplitude to vanish identically.

Microscopic analysis of the two-particle propagator and its relation to the quasiparticle 
scattering amplitude leads to an identity between the scattering amplitude in the forward 
direction, and the Landau parameters, $F_{\ell}^{s,a}$, that define the quasiparticle 
molecular fields. In terms of the symmetric and anti-symmetric amplitudes in the p-h 
channel, Landau's identity for the forward scattering amplitude is\cite{lan59},
\be
\label{forward_scattering}
\Tm^{s,a}(\theta,\phi=0) = \sum_{\ell=0}^{\infty}\,A^{s,a}_{\ell}\,\cP_{\ell}(\cos\theta)
\,,
\ee
where
\be
A^{s,a}_{\ell} = \frac{F^{s,a}_{\ell}}{1+F^{s,a}_{\ell}/(2\ell+1)}
\,.
\ee
In terms of the singlet and triplet amplitudes for $\phi=0$,
\ber
\label{singlet_forward_scattering}
\Tm_{s}(\theta,\phi=0) 
&=& 
\sum_{\ell\ge 0}\left(A^{s}_{\ell}-3 A^{a}_{\ell}\right)\cP_{\ell}(\cos\theta)
\\
\Tm_{t}(\theta,\phi=0) 
&=& 
\sum_{\ell\ge 0}\left(A^{s}_{\ell}+A^{a}_{\ell}\right)\cP_{\ell}(\cos\theta)
\,.
\label{triplet_forward_scattering}
\eer

Exchange symmetry leads to an additional constraint on the triplet amplitude. 
In the limit $\theta=0$, $\phi=0$, i.e. for $\hat\vp_1=\hat\vp_2=\hat\vp_3=\hat\vp_4$, 
the triplet amplitude necessarily vanishes. Thus, from 
Eq. (\ref{triplet_forward_scattering}) we obtain the forward scattering 
sum rule (FSSR),
\be\label{FSSR}
\lim_{\theta\rightarrow 0}\Tm_{t}(\theta,0)
=
\sum_{\ell\ge 0}\left(A^{s}_{\ell}+A^{a}_{\ell}\right)
\equiv 0
\,.
\ee

\subsubsection*{s-p-d Scattering}

Several microscopic and phenomenological theories have been proposed for the 
quasiparticle scattering amplitude in \He.\cite{dy69,sau81b,pfi83,lev83}
Here we adopt a slightly modified version of the model proposed by Dy and Pethick.\cite{dy69} 
They proposed a minimal model for the scattering amplitude that obeys exchange 
anti-symmetry. In particular, if we assume the singlet and triplet scattering 
amplitudes are to a good approximation given by the $m=0$ and $m=1$ terms, we have 
$\Tm_{s}\simeq A_{s}(\cos\theta)$ and $\Tm_{t}\simeq A_{t}(\cos\theta)\cos\phi$. 
In this case we can fix the expansion coefficients of $A_{t,s}(\cos\theta)$ in terms 
of the forward-scattering amplitudes, $A_{\ell}^{s,a}$, and thus the Landau 
parameters, $F_{\ell}^{s,a}$,\cite{dy69}
\ber
&
\label{singlet_s-wave}
\hspace{-10mm}
\Tm_{s}\simeq& 
       \sum_{\ell\ge 0}\left(A^{s}_{\ell}-3A^{a}_{\ell}\right)
       \,\cP_{\ell}(\cos\theta)
\,,
\\
&
\label{triplet_p-wave}
\hspace{-10mm}
\Tm_{t}\simeq& 
       \sum_{\ell\ge 0}\left(A^{s}_{\ell}+A^{a}_{\ell}\right)
       \,\cP_{\ell}(\cos\theta)\,\cos\phi
\,.
\eer

The quasiparticle lifetime, $\tau_{\text{in}}(T)$ in Eq. \ref{tau_in}, 
as well as the thermal transport time, $\tau_{\kappa}(T)$, due to binary 
quasiparticle collisions in pure \He\ are determined by angular averages 
of the spin-averaged transition probability,
\ber
\frac{1}{\tau_{\text{in}}}&=&\frac{N_f^2}{v_f p_f}\,
			     \langle W\rangle\,
                             \left(k_B T\right)^2 
\\
\tau_{\kappa}             &=& S^{\infty}_{\kappa}(\lambda_{\kappa})
                              \tau_{\text{in}}
\,,
\eer
with
\be
\lambda_{\kappa}\equiv\langle W\left(1+2\cos\theta\right)\rangle/\langle W \rangle
\,,
\ee
where $W=\onefourth W_{\uparrow\uparrow}+\onehalf W_{\uparrow\downarrow}$ and 
the angular average is defined in Eq. \ref{Fermi-surface_average}. 
Writing $W_{ab} = \frac{\pi}{2}\hbar^{-1}N_f^{-2}\,\bar{W}_{ab}$, the 
transition probability can be expressed in terms of the  
dimensionless singlet and triplet scattering amplitudes,
\be\label{dimensionless_rate}
\bar{W}=\left|\Tm_s\right|^2+3\left|\Tm_t\right|^2 
       + 2\Re\left\{\Tm_{t}\Tm_{s}^*\right\}
\,,
\ee
and the quasiparticle lifetime becomes,
\be\label{tau_in-T2}
\frac{1}{\tau_{\text{in}}}=\frac{\pi}{32}\,\hbar^{-1}
                           \frac{\left(k_B T\right)^2}{E_f}\, 
			   \langle\bar{W}\rangle
\,.
\ee
Note that for weighted averages of $\bar{W}$ in which the weight
function is even under exchange (i.e. $\phi\rightarrow\phi+\pi$) the
cross term in Eq. (\ref{dimensionless_rate}) vanishes.
Similarly, for the thermal transport time we can write 
$\lambda_{\kappa}=\Lambda_{\kappa}/\langle \bar{W}\rangle$ where
\be
\Lambda_{\kappa}=\langle\left(1+2\cos\theta\right)\bar{W}\rangle
\,.
\ee

In the {\sl spd} model the Fermi-surface average of the rate 
becomes,
\be\label{W}
\langle \bar{W}\rangle 
= 
\langle A_s^2\rangle 
+
\threehalves\langle A_t^2\rangle
\,,
\ee
We evaluate this rate in terms of Legendre expansion of the 
forward-scattering amplitudes. For either singlet or triplet channel,
\be
A(\cos\theta)=\sum_{\ell}A_{\ell}\,P_{\ell}(\cos\theta)
\,.
\ee
The angular average of $A^2$ is given by
\ber
\langle A^2\rangle 
&=&\sum_{\ell\ell'}\,C_{\ell\ell'}\,A_{\ell}A_{\ell'}
\,,
\eer
where 
\be\label{C_coefficients}
C_{\ell\ell'}\equiv \int_{0}^{1} dx\,
 P_{\ell}(2x^2-1)\,P_{\ell'}(2x^2-1)
\,.
\ee 
Similarly, for the angular averages of the form,
\be
\langle A^2\left(1+2\cos\theta\right)\rangle =  
\sum_{\ell\ell'}\,L_{\ell\ell'}\,A_{\ell}A_{\ell'}
\,,
\ee
with
\be\label{L_coefficients}
L_{\ell\ell'}\equiv \int_{0}^{1} dx\,
\left(4x^2-1\right)P_{\ell}(2x^2-1)\,P_{\ell'}(2x^2-1)
\,.
\ee
These coefficients are listed in Table \ref{table-CL_coefficients}
for $\ell, \ell' \le 2$. 

\begin{table}[ht]
\begin{center}
\begin{tabular}{rl}
{\small
\begin{tabular}{|c||c|c|c|}
\hline
$C_{\ell\ell'}$		& $0$		& $1$		& $2$ 		\\
\hline
\hline										 
$0$			& $1$		& $-1/3$	& $1/5$		\\
\hline										 
$1$			& $-1/3$	& $7/15$	& $-23/105$	\\
\hline										 
$2$			& $1/5$		& $-23/105$	& $11/105$	\\
\hline
\end{tabular}
}
&
{\small
\begin{tabular}{|c||c|c|c|}
\hline
$L_{\ell\ell'}$		& $0$		& $1$		& $2$ 		\\
\hline
\hline										 
$0$			& $1/3$		& $3/5$		& $-5/21$	\\
\hline										 
$1$			& $3/5$		& $-1/21$	& $1/3$		\\
\hline										 
$2$			& $-5/21$	& $1/3$		& $-71/1155$	\\
\hline
\end{tabular}
}
\end{tabular}
\end{center}
\caption{Coefficients defining the angular averages of $\langle \bar{W}\rangle$ 
         ($C_{\ell\ell'}$) and $\langle \bar{W}(1+2\cos\theta)\rangle$ 
	 ($L_{\ell\ell'}$) in the spd scattering model.}
\label{table-CL_coefficients}
\end{table}

The input for our calculations of the transport properties of bulk \He\ as
well as \Heaero\ are the Fermi-liquid parameters. The measured values of these
parameters are collected in Ref. \onlinecite{hal90}, and are also available 
online.\cite{har00}
The Landau interaction parameters, $F_{0}^{s}$, $F_{1}^{s}$, and $F_{0}^{a}$ 
are accurately known from measurements of the heat capacity, first-sound velocity 
and magnetic susceptibility of pure normal \He, while determinations of 
$F_{2}^{s}$ and $F_{1}^{a}$ have also been obtained from measurements of the 
zero sound velocity and spin-wave resonance for normal \He, respectively. 
However, these parameters are not as accurately determined. 
Less is known about the magnitude and pressure dependence of the $\ell =2$ 
contribution to the exchange interaction, $F_2^a$,\cite{fis86} and much less
is known quantitatively about the Landau interaction parameters corresponding 
to harmonics $\ell > 2$, although evidence of interactions in higher
order scattering channels is suggested by the observation of 
high frequency pair exciton\cite{sau81} modes in superfluid \Heb.\cite{dav08}

The scattering model we use throughout is defined by Eqs. (\ref{singlet_s-wave}) 
and (\ref{triplet_p-wave}) with the added assumption that we truncate the expansion, 
i.e. set $A_{\ell}^{s,a}=0$ for $\ell \ge 3$. This approximation is reasonable if the 
contributions to the Fermi-surface averages of the scattering rate fall off sufficiently 
rapidly with increasing $\ell >2$.

The {\sl spd} model with the Fermi liquid data for $\ell \le 2$ as input qualitatively describes the
decrease in the transport time, $\tau_{\kappa}T^2$, with increasing pressure (shown in Fig.
\ref{fig-bulk_thermal_transport_time}) and is within 25\% of the experimental values for
$\tau_{\kappa}T^2$ over the full pressure range.
However, the comparison clearly shows that the {\sl spd} model, or the accuracy of 
the known Fermi liquid data is inadequate, or both.
The most problematic aspect of the {\sl spd} model as it stands is that the  
FSSR is badly violated, when evaluated with $A^{s,a}_{\ell}=0$ for $\ell \ge 3$.
In particular the largest violation in the FSSR, 
\be \label{FFSR-error}
S_{\tiny{\sf error}}  = \sum_{\ell}\left(A^s_{\ell} + A^a_{\ell}\right) \approx -1.0
\,,
\ee
is at low pressures, which is also where the discrepancy (refer to 
Fig. \ref{fig-bulk_thermal_transport_time})  
between theory (solid line) and experiment (red diamonds) is greatest.
This is a significant violation of the Pauli exclusion principle, and is an 
indication that either the determinations of $F^{a}_{1,2}$ are inaccurate, 
that there is significant weight in the interaction channels for $\ell \ge 3$, 
or both. 

Respecting the Pauli exclusion principle, by enforcing the FSSR, is likely
more important than knowing precisely the distribution of higher angular momentum 
channels that account for the missing weight in Eq. \ref{FFSR-error}. 
Thus, we enforce the FSSR by fixing the least known material parameter in the 
{\sl spd} model, i.e. we replace
\be
A_2^a \rightarrow A_2^{a'} = A_2^{a} -S_{\tiny{\sf error}}
\,.
\ee
The importance of enforcing the FSSR appears to be born out by the improvement between theory (black
dots) and experiment shown in Fig. \ref{fig-bulk_thermal_transport_time}. For pressures below $p\le
25\,\mbox{bar}$ the agreement is nearly perfect. Thus, the deviations between theory and experiment
at higher pressures likely reflects real limitations of the {\sl spd} model, i.e. there is
scattering that reduces heat transport that is outside the \emph{spd} scattering model.


\end{document}